\documentclass[graybox]{svmult}

\usepackage{mathptmx}       
\usepackage{helvet}         
\usepackage{courier}        
\usepackage{type1cm}        
\usepackage{makeidx}         
\usepackage{graphicx}        
\usepackage{multicol}        
\usepackage[bottom]{footmisc}
\usepackage{amsmath}
\usepackage{amssymb}

\makeindex             

\begin{document}
  \newcommand {\nc} {\newcommand}
  \nc {\beq} {\begin{eqnarray}}
  \nc {\eeq} {\nonumber \end{eqnarray}}
  \nc {\eeqn}[1] {\label {#1} \end{eqnarray}}
  \nc {\eol} {\nonumber \\}
  \nc {\eoln}[1] {\label {#1} \\}
  \nc {\ve} [1] {\vec{#1}}
  \nc {\ves} [1] {\mbox{\boldmath ${\scriptstyle #1}$}}
  \nc {\mrm} [1] {\mathrm{#1}}
  \nc {\half} {\mbox{$\frac{1}{2}$}}
  \nc {\thal} {\mbox{$\frac{3}{2}$}}
  \nc {\fial} {\mbox{$\frac{5}{2}$}}
  \nc {\la} {\mbox{$\langle$}}
  \nc {\ra} {\mbox{$\rangle$}}
  \nc {\etal} {{\it et al.}}
  \nc {\eq} [1] {(\ref{#1})}
  \nc {\Eq} [1] {Eq.~(\ref{#1})}
  \nc {\Ref} [1] {Ref.~\cite{#1}}
  \nc {\Refc} [2] {Refs.~\cite[#1]{#2}}
  \nc {\Sec} [1] {Sec.~\ref{#1}}
  \nc {\chap} [1] {Chapter~\ref{#1}}
  \nc {\anx} [1] {Appendix~\ref{#1}}
  \nc {\tbl} [1] {Table~\ref{#1}}
  \nc {\Fig} [1] {Fig.~\ref{#1}}
  \nc {\ex} [1] {$^{#1}$}
  \nc {\Sch} {Schr\"odinger }
  \nc {\flim} [2] {\mathop{\longrightarrow}\limits_{{#1}\rightarrow{#2}}}
  \nc {\textdegr}{$^{\circ}$}
  \nc {\inred} [1]{\textcolor{red}{#1}}
  \nc {\inblue} [1]{\textcolor{blue}{#1}}
  \nc {\IR} [1]{\textcolor{red}{#1}}
  \nc {\IB} [1]{\textcolor{blue}{#1}}
  \nc{\pderiv}[2]{\cfrac{\partial #1}{\partial #2}}
  \nc{\deriv}[2]{\cfrac{d#1}{d#2}}
  \nc {\R} {\mathbb{R}}

\title*{Introduction to Nuclear-Reaction Theory}
\author{Pierre Capel}
\institute{Pierre Capel \at Institut f\"ur Kernphysik, Johannes Gutenberg-Universit\"at Mainz, D-55099 Mainz, Germany \& Physique Nucl\'eaire et Physique Quantique (CP 229),
Universit\'e libre de Bruxelles (ULB), B-1050 Brussels,\\ \email{pcapel@uni-mainz.de}}
%
%
\maketitle

\abstract*{These notes summarise the lectures I gave during the summer school ``International Scientific Meeting on Nuclear Physics'' at La R\'abida in Spain in June 2018.
They offer an introduction to nuclear-reaction theory, starting with the basics in quantum scattering theory followed by the main models used to describe breakup reactions: the Continuum Discretised Coupled Channel method (CDCC), the Time-Dependent approach (TD) and the eikonal approximation.
These models are illustrated on the study of the exotic structure of halo nuclei.}

\abstract{These notes summarise the lectures I gave during the summer school ``International Scientific Meeting on Nuclear Physics'' at La R\'abida in Spain in June 2018.
They offer an introduction to nuclear-reaction theory, starting with the basics in quantum scattering theory followed by the main models used to describe breakup reactions: the Continuum Discretised Coupled Channel method (CDCC), the Time-Dependent approach (TD) and the eikonal approximation.
These models are illustrated on the study of the exotic structure of halo nuclei.}

\section*{\label{intro}Introduction}
Nuclear reactions are used for a variety of goals.
They can be used to study the structure of nuclei; sometimes, they can be the only way to probe nuclear structure, especially far from stability.
Nuclear reactions also provide information about the interaction between nuclei, either to study the fundamentals of the nuclear force, or to measure reaction rates, which are major inputs in other fields of physics, like nuclear astrophysics, or in a broad range of nuclear applications, like nuclear power or the production of radioactive isotopes for medical purposes.

To correctly analyse and exploit data of reaction measurements, it is important to know the basics in nuclear-reaction theory.
The present notes offer an introduction to this exciting discipline.
\Sec{scatt} presents the basics of non-relativistic scattering theory for two colliding particles, which interact through a potential.
In this section, the notion of \emph{cross section} is introduced and its calculation from the solution of the stationary \Sch equation is explained.
In particular, the method based on the \emph{partial-wave} expansion of the wave function is presented in detail.
To close this first chapter, I introduce the \emph{optical model}, which enables to account for other reaction channels that can take place during the collision of the particles.
In these developments, I closely follow the Chapter~VIII of the textbook on quantum mechanics by Cohen-Tannoudji, Diu and Lalo\"e \cite{CT77}.
For interested readers, a more detailed presentation of quantum reaction theory can be found in \Ref{Tay72}.

In \Sec{bu}, I give a brief overview of the main methods used to describe \emph{breakup} reactions.
That sections starts with a presentation of \emph{halo nuclei}, which are one of the most exotic quantal structures found far from stability, and which are studied mostly through reactions, like breakup.
I then pursue with the three main models of breakup: the Continuum Discretised Coupled Channel method (CDCC), the Time-Dependent approach (TD) and the eikonal approximation.
The section closes with a comparison between them that emphasises the advantage and drawbacks of each of these models, and which gives their respective range of validity.
More advanced developments on nuclear-reaction theory can be found in Refs.~\cite{NT09,BD04}.

In \Sec{structure}, I review the information about the structure of halo nuclei that can be inferred from the analysis of breakup measurements.
We will see in this section what can be expected from experimental data, and, most importantly, what cannot be inferred from experiments.
This section is built mostly from recent articles published in the literature.
Their selection of course reflects my personal biases on the subject as well as my own research activity.

I do not believe this paper exhausts the vast subject of nuclear-reaction theory, but I hope that it will give an incentive to some of the readers to pursue their journey in the landscape of nuclear physics within this exciting and flourishing field of research.
Without further ado, let us start this introduction with the basics in quantum collision theory.

\section{\label{scatt}Quantum Collision Theory}


\subsection{\label{types}Types of Collisions}

Quantum collisions are used in various applications.
They are sometimes one of the only way to study the interaction between particles.
For example, the potentials used in nuclear-structure calculations to simulate the interaction between the nucleons is deduced mostly from observables measured in nucleon-nucleon collisions \cite{NAR13} (see \Sec{phaseshifts}).
Collisions are also used to infer information about the structure of quantal objects.
The famous experiment of Rutherford, Geiger and Marsden performed in 1909 is a good example. 
This experiment, in which alpha particles were fired at a gold foil, enabled the discovery of the structure of atoms.
Nowadays, reactions are measured to study the structure of nuclei throughout the whole nuclear chart.
In a more natural way, collisions can also be used to obtain reaction rates of particular interest, e.g., for reactions that take place in stars or that are needed in technological applications, like nuclear reactors or to produce radioactive isotopes of medical use.

\begin{figure}[t]
\center
\includegraphics[width=\linewidth]{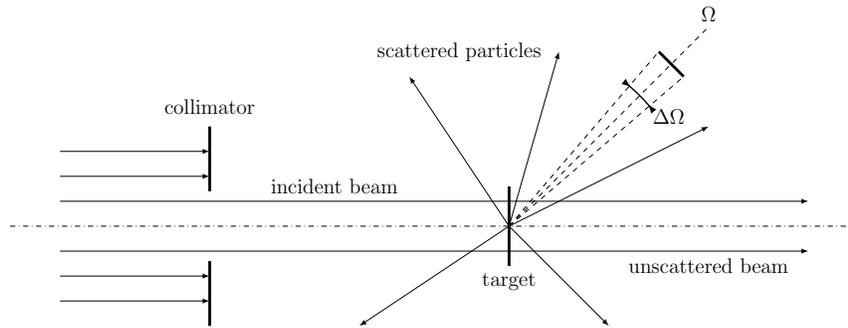}
\caption{\label{f1} Typical measurement scheme of nuclear reactions.}
\end{figure}

The usual measurement scheme of these reactions in nuclear physics is schematically illustrated in \Fig{f1}.
An incident beam made up of projectile particles, coming from the left, is first collimated before impinging on a fixed target.
The particles produced during the collision are then scattered in all possible directions $\Omega\equiv(\theta,\varphi)$ and measured in the detectors surrounding the target.
These detectors have a finite size and are seen from the target under a solid angle $\Delta\Omega$.
Note that a significant amount of the incoming particles will not react and continue straight ahead, undeflected by the target.
Due to this unscattered beam, measuring reactions at very small scattering angle can be quite difficult, when not impossible.

When two quantal objects collide, various reactions can take place.
Let us consider the collision of two ``particles'' $a$ and $b$, where the term particle is taken in a broad sense and can mean molecules, atoms, nuclei, nucleons,\ldots
\begin{enumerate}
\item $a+b\rightarrow a+b$ \hspace{8.5mm}(elastic scattering)
\item \hspace{8mm}$\rightarrow a+b^*$ \hspace{7mm}(inelastic scattering)
\item \hspace{8mm}$\rightarrow c+f+b$ \hspace{3mm}(breakup)
\item \hspace{8mm}$\rightarrow d+e$ \hspace{8.5mm}(rearrangement or transfer)
\end{enumerate}
The first and most evident one is the \emph{elastic scattering}, in which the two particles merely scatter while remaining in their initial states.
Second, if the incident energy is high enough, some energy can be transferred from the relative motion of $a$ and $b$ towards one of the---or both---particles, which then leave(s) the collision in an excited state.
Such a state is denoted by the asterisk ($^*$) next to the excited particle.
The scattering is then said \emph{inelastic}.
A third possibility is that this energy transfer is high enough to \emph{break up} one of the particles into its more elementary constituents (e.g. in two ions in the case of a collision of molecules, into an ion and an electron if the collision involves atoms, or into nucleon clusters in a nuclear collision).
The fourth case envisaged here is the case of \emph{rearrangement} or \emph{transfer}, in which some sub-particles are transferred from the projectile to the target or vice versa (like atoms in molecular collisions, electrons in atomic collisions, or nucleons or groups of nucleons in nuclear reactions).

As an example in the realm of nuclear physics, let us consider the collision of $^{11}$Be on $^{208}$Pb, which can be \ref{Be1}. elastically scattered, \ref{Be2}. inelastically scattered (e.g. if $^{11}$Be is excited to its $\half^-$ state during the collision), \ref{Be3}. broken up into a $^{10}$Be and a neutron (see \Sec{bu}), or \ref{Be4}. have that valence neutron transferred to the $^{208}$Pb target to form $^{209}$Pb:
\begin{enumerate}
\item \label{Be1} $\rm ^{11}Be+\,^{208}Pb\rightarrow \,^{11}Be+\,^{208}Pb$ \hspace{18mm}(elastic scattering)
\item \label{Be2} \hspace{19.5mm}$\rightarrow \rm \,^{11}Be^*(1/2^-)+\,^{208}Pb$ \hspace{6mm}(inelastic scattering)
\item \label{Be3} \hspace{19.5mm}$\rightarrow \rm \,^{10}Be+n+\,^{208}Pb$ \hspace{12mm}(breakup)
\item \label{Be4} \hspace{19.5mm}$\rightarrow \rm \,^{10}Be+\,^{209}Pb$ \hspace{17.5mm}(transfer)
\end{enumerate}

Although the reactions taking place during the collision can significantly affect the structure of the colliding particles, some physical values are conserved.
One particular case is the total energy of the system, which remains the same before and after the collision.
Its conservation means that the sum of the mass energy and total kinetic energy in any outgoing channel must equal that in the incoming channel:
\beq
m_a\, c^2+m_b\, c^2+T_{\rm in}=\sum_i m_i\, c^2 + T_{\rm out}
\eeqn{e1}
where $m_a$ and $m_b$ are the masses of the colliding particles $a$ and $b$, respectively, and $T_{\rm in}$ is the total kinetic energy in the incoming channel.
In the right-hand side of \Eq{e1} $m_i$ are the masses of all the particles produced during the collision in one particular channel, and $T_{\rm out}$ is their total kinetic energy.

From this expression, one can define the \emph{$Q$ value} of a particular reaction, which corresponds to the energy ``produced'' by this reaction
\beq
Q=m_a\, c^2+m_b\, c^2-\sum_i m_i\, c^2.
\eeqn{e2}
If $Q>0$ the reaction is said \emph{exoenergetic} as it \emph{produces energy}.
Energetically, the reaction is then always possible.
On the contrary, if $Q<0$, the reaction is said \emph{endoenergetic} and requires a minimal initial kinetic energy to take place.
On the sole energy viewpoint, a channel will be said \emph{open} if $T_{\rm in}>-Q$.
If this is not the case, i.e. if the reaction is endoenergetic and if the incident kinetic energy is lower than $|Q|$, the channel is said \emph{closed}.
Note that the elastic-scattering channel is always open, since in that case $Q=0$.

\clearpage

\subsection{Notion of Cross Section}
To characterise a reaction and measure its probability to take place during the collision of two particles, we use the notion of \emph{cross section}.
To introduce this observable, let us go back to the measurement scheme pictured in \Fig{f1}.
The number of particles $\Delta n$ detected in the direction $\Omega$ per unit time within the solid angle $\Delta\Omega$ covered by the detector will be naturally proportional to the flux $F_i$ of incoming particles and the number $N$ of particles within the target:
\beq
\Delta n=F_i\ N\ \Delta\sigma.
\eeqn{e3}
the factor of proportionality $\Delta \sigma$ is the \emph{cross section} for the outgoing channel considered in the measurement.
The dimensions of $\Delta n$ are that of a number of event per unit time, those of $F_i$ that of a number of particles per unit time and unit area, and $N$ is of course just a number of particles.
Consequently, the dimensions of $\Delta\sigma$ are that of an area.
Its units are usually expressed in \emph{barns} (b) : 1~b~$=10^{-24}$~cm$^2=100$~fm$^2$.

The \emph{differential cross section} corresponds to the limit
\beq
\frac{d\sigma}{d\Omega}&=&\lim_{\Delta\Omega\rightarrow0}\frac{\Delta\sigma}{\Delta \Omega}\label{e4a}\\
 &=&\lim_{\Delta\Omega\rightarrow0}\frac{\Delta n}{F_i N\Delta\Omega}.
\eeqn{e4}
The direction $\ve{\hat Z}$ is often chosen along the beam axis.
In that case, thanks to the cylindrical symmetry of the problem, the cross section $d\sigma/d\Omega$ depends only on the colatitude $\theta$ and is independent of the azimuthal angle $\varphi$.

This schematically explains how these values can be measured.
To see how they can be calculated, let us consider the elastic scattering of two particles, $a$ and $b$, which interact through a potential $V$.
This potential depends on the $a$-$b$ relative coordinate $\ve{R}=\ve{R}_b-\ve{R}_a$ (for the sake of clarity, the spin of the particles is neglected in this development).
The Hamiltonian of the system hence reads
\beq
{\cal H}(\ve{R}_a,\ve{R}_b)=T_a+T_b+V(\ve{R}),
\eeqn{e5}
where the kinetic energy of particles $a$ and $b$, respectively, read
\beq
T_a&=&\frac{p_a^2}{2m_a}=-\frac{\hbar^2\Delta_{R_a}}{2m_a} \label{e6}\\
T_b&=&\frac{p_b^2}{2m_b}=-\frac{\hbar^2\Delta_{R_b}}{2m_b},
\eeqn{e7}
with $\ve{p}_a$ and $\ve{p}_b$ the momenta of $a$ and $b$, respectively.

Since $V$ depends only on the relative coordinate $\ve{R}$ it is useful to change coordinates and use, instead of $\ve{R}_a$ and $\ve{R}_b$, their relative coordinate $\ve{R}$ and the coordinate of their centre of mass
\beq
\ve{R}_{\rm cm}=\frac{m_a\ve{R}_a+m_b\ve{R}_b}{M},
\eeqn{e8}
were $M=m_a+m_b$ is the total mass in the incoming channel.
Within this new set of coordinates, the Hamiltonian of the system reads
\beq
{\cal H}(\ve{R}_a,\ve{R}_b)=T_{\rm cm}+T_{R}+V(\ve{R}),
\eeqn{e10}
where
\beq
T_{\rm cm}=\frac{P_{\rm cm}^2}{2M}=-\frac{\hbar^2\Delta_{R_{\rm cm}}}{2M}
\eeqn{e11}
is the kinetic energy of the centre of mass of $a$ and $b$, with $\ve{P}_{\rm cm}$ its momentum, and
\beq
T_R=\frac{P^2}{2\mu}=-\frac{\hbar^2\Delta_{R}}{2\mu},
\eeqn{e12}
is the kinetic energy of the relative motion between $a$ and $b$, with $\ve{P}$ the corresponding momentum and $\mu=m_a m_b/M$ the \emph{reduced mass} of $a$ and $b$.

It follows from \Eq{e10} that $\cal H$ is the sum of two Hamiltonians, which are functions of two independent variables $\ve{R}_{\rm cm}$ and $\ve{R}$:
\beq
{\cal H}(\ve{R}_a,\ve{R}_b)=H_{\rm cm}(\ve{R}_{\rm cm})+H(\ve{R}).
\eeqn{e13}
Accordingly the wave function that describes this two-particle system can be factorised into
\beq
\Psi_{\rm tot}(\ve{R}_a,\ve{R}_b)=\Psi_{\rm cm}(\ve{R}_{\rm cm})\ \Psi(\ve{R}).
\eeqn{e14}

The wave function $\Psi_{\rm cm}$ describes the motion of the centre of mass of $a$ and $b$.
It is solution of the \Sch equation
\beq
H_{\rm cm}\ \Psi_{\rm cm}(\ve{R}_{\rm cm})&=&E_{\rm cm}\ \Psi_{\rm cm}(\ve{R}_{\rm cm}),
\eeqn{e15}
where $H_{\rm cm}=T_{\rm cm}$ [see \Eq{e10}], which describes the motion of a free particle of mass $M$.
For a particle of initial momentum $\ve{P}_{\rm cm}=\hbar \ve{K}_{\rm cm}$, $\Psi_{\rm cm}$ corresponds simply to a plane wave
\beq
\Psi_{\ve{K}_{\rm cm}}(\ve{R}_{\rm cm})=(2\pi)^{-3/2}e^{i\ve{K}_{\rm cm}\cdot\ve{R}_{\rm cm}},
\eeqn{e16}
with $\hbar^2K^2_{\rm cm}/2M=E_{\rm cm}$, the kinetic energy of the $a$-$b$ centre of mass in the reference frame in which $\ve{R}_a$ and $\ve{R}_b$ are defined.
The normalisation factor $(2\pi)^{-3/2}$ is chosen such that
\beq
\langle \Psi_{\ve{K'}_{\rm cm}}|\Psi_{\ve{K}_{\rm cm}}\rangle=\delta(\ve{K}_{\rm cm}-\ve{K'}_{\rm cm})
\eeqn{e17}

This wave function \eq{e16} hence describes a centre of mass in uniform translation, as we would have expected from Galilean invariance.
Accordingly, this motion does not add anything and can thus be ignored.
The physics of the problem is thus entirely captured within the Hamiltonian
\beq
H(\ve{R})=T_R+V(\ve{R}),
\eeqn{e18}
which describes the relative motion of particles $a$ and $b$.
In scattering theory, the meaningful eigenstates of $H$ are the \emph{stationary scattering states}.

\subsection{Stationary Scattering States}
A stationary scattering state $\Psi_{K\ve{\hat Z}}$ is a solution of
\beq
H\ \Psi_{K\ve{\hat Z}}(\ve{R})=E\ \Psi_{K\ve{\hat Z}},
\eeqn{e19}
which exhibits the following asymptotic behaviour
\beq
\Psi_{K\ve{\hat Z}}(\ve{R})\flim{R}{\infty}(2\pi)^{-3/2}\left[e^{iKZ}+f_K(\theta)\,\frac{e^{iKR}}{R}\right].
\eeqn{e20}
In Eqs.~\eq{e19} and \eq{e20}, $\ve{\hat Z}$ has been chosen as the beam axis, for which choice the expression does not depend on the azimuthal angle $\varphi$ as explained above.

The solutions of \Eq{e19} we are looking for behave asymptotically as the sum of a plane wave $e^{iKZ}$ and an outgoing spherical wave $f_K(\theta){e^{iKR}}/{R}$, whose amplitude is modulated as a function of the scattering angle $\theta$ by the function $f_K$, which is called the \emph{scattering amplitude}.
Note that, for both terms in \Eq{e20}, the momentum $\hbar K$ is related to the energy $E=\hbar K^2/2\mu$.
To interpret the physical meaning of this asymptotic behaviour, let us recall the operator of the current of probability
\beq
\ve{J}(\ve{R})=\frac{1}{\mu}\Re[\Psi^*(\ve{R})\ \ve{P}\ \Psi(\ve{R})],
\eeqn{e21}
and let us compute this operator on each term of \Eq{e20}.
For the plane wave, we obtain
\beq
\ve{J}_i(\ve{R})=(2\pi)^{-3}\frac{\hbar K}{\mu}\ve{\hat Z}=(2\pi)^{-3}\,v\ \ve{\hat Z}
\eeqn{e22}
where $v=\hbar K/\mu$ is the incoming $a$-$b$ relative velocity.
We can thus interpret this term as the incoming current of probability, describing the projectile impinging on the target at a velocity $v$ along the beam axis.

For the spherical wave, we get
\beq
\ve{J}_s(\ve{R})=(2\pi)^{-3}\,v\ |f_K(\theta)|^2\frac{1}{R^2}\ve{\hat R}+{\cal O}\left(\frac{1}{R^3}\right),
\eeqn{e23}
which, at large distance, is purely radial and directed outwards.
This current is still proportional to $v$, but its magnitude varies with $\theta$ according to the square modulus of the scattering amplitude $|f_K(\theta)|^2$.
It can thus be seen as the scattered current that describes the relative motion of the two particles after they have interacted.

To obtain a formal expression of the cross section, let us go back to its definition \eq{e4}.
Following the physical interpretation of the asymptotic behaviour of the scattering state $\Psi_{K\ve{\hat K}}$, we can assume that in a quantal description of the process, the incoming flux $F_i$ [see \Eq{e3}] will be proportional to the incoming current $J_i$ \eq{e22}
\beq
F_i=C\ J_i.
\eeqn{e24}
In the same line of thought, the flux of particle scattered in direction $\Omega$ is related to the scattered current $J_s$ \eq{e23} by
\beq
F_s=C\ J_s.
\eeqn{e25}
For a single scattering centre, i.e. $N=1$, the number of particles $\Delta n$ observed in a given direction per unit time in a detector of section $\Delta S$ will thus be
\beq
\Delta_n&=&F_s\ \Delta S\\
 &=&C\ J_s\ R^2\ \Delta\Omega,
\eeqn{e26}
where $\Delta\Omega$ is the solid angle under which the detector is seen from the target when it is placed at a distance $R$.
Following the definition \eq{e4a}, the differential cross section for the scattering of $a$ by $b$ hence reads
\beq
\frac{d\sigma}{d\Omega}&=&\lim_{\Delta\Omega\rightarrow0}\frac{\Delta n}{F_i\,\Delta\Omega}\\
 &=&\frac{R^2J_s}{J_i}\label{e28}\\
 &=&|f_K(\theta)|^2.
\eeqn{e29}
The cross section is therefore just the square modulus of the scattering amplitude.
This means that all the effect of the interaction between $a$ and $b$---viz. of the potential $V$---on the scattering process is included in $f_K$, as expected from the physical interpretation we gave of $J_s$ \eq{e23}.
In order to obtain that scattering amplitude, it is necessary to solve the \Sch equation \eq{e19} under the asymptotic condition \eq{e20}.
There exist various techniques to do that.
In the next section, we will see one that works for central potentials and that is often used for low-energy scattering: the partial-wave expansion.

\subsection{\label{phaseshifts}Partial-Wave Expansion and Phasesift}

\subsubsection{Partial-Wave Expansion}

When the potential is central, i.e. when it depends only on the distance $R$ between $a$ and $b$, the Hamiltonian $H$ commutes with the orbital angular momentum operators ${\cal L}^2$ and ${\cal L}_Z$.
The wave function $\Psi_{K\ve{\hat{Z}}}$, solution of \Eq{e19}, which describes the $a$-$b$ relative motion can then be expanded upon the eigenfunctions of ${\cal L}^2$ and ${\cal L}_Z$, which are the \emph{spherical harmonics} $Y_L^M$
\beq
\Psi_{K\ve{\hat Z}}(\ve{R})=(2\pi)^{-3/2}\frac{1}{KR}\sum_{L=0}^{\infty}c_L\ u_{KL}(R)\ Y_L^0(\theta),
\eeqn{e30}
taking into account the aforementioned cylindrical symmetry.
Including this expansion in \Eq{e19}, we obtain the following equation for the reduced radial wave functions $u_{KL}$
\beq
\left(\frac{d^2}{dR^2}-\frac{L(L+1)}{R^2}-\frac{2\mu}{\hbar^2}V(R)+K^2\right)u_{KL}(R)=0.
\eeqn{e33}
This equation can be solved using numerical techniques.
One advantage of this decomposition is to reduce the three-dimensional problem \eq{e19} to one dimension.
Of course, \Eq{e33} will have to be solved for all the values of $L$.
However, as we will see later, especially at low energy $E$, the sum over $L$ in the expansion \eq{e30} can be truncated to a limited number of terms.

\subsubsection{\label{phaseshift}Phaseshift}

For now, let us assume that \Eq{e33} can be solved for all the values of $L$ that are needed.
Our goal being to calculate the cross section \eq{e29}, we need to evaluate the scattering amplitude $f_K$, which appears in the asymptotic expression of the stationary scattering states \eq{e20}.
To see how to obtain $f_K$ from the solution of \Eq{e33}, let us study the behaviour of $u_{KL}$ at large $R$.
If we assume the interaction potential $V$ to be short-ranged, viz. $R^2\,V(R)\flim{R}{\infty}0$, the asymptotic solution of \Eq{e33} $u_{KL}^{\rm as}$ is solution of
\beq
\left(\frac{d^2}{dR^2}-\frac{L(L+1)}{R^2}+K^2\right)u_{KL}^{\rm as}(R)=0,
\eeqn{e34}
whose solutions are known analytically and exhibit a well-known asymptotic behaviour
\beq
u_{KL}^{\rm as}(R)&=&A\ KR\ j_L(KR)+B\ KR\ n_L(KR)\\
 &\flim{R}{\infty}&A\, \sin(KR-L\pi/2)+B\, \cos(KR-L\pi/2)
\eeqn{e35}
where $j_L$ and $n_L$ are the spherical Bessel functions of the first and second kinds, respectively.
We can re-define the constants $A$ and $B$ as $A=C\,\cos \delta_L$ and $B=C\,\sin\delta_L$, respectively.
This gives us
\beq
u_{KL}^{\rm as}(R)\flim{R}{\infty}C\, \sin(KR-L\pi/2+\delta_L),
\eeqn{e35}
where $C$ is just an overall normalisation constant, whereas $\delta_L$, which is called the \emph{phaseshift}, contains all the information about the scattering potential.
To better grasp the physical meaning of this phaseshift, let us re-write \Eq{e35} as the sum of an incoming and an outgoing spherical waves
\beq
u_{KL}(R)&\flim{R}{\infty}&i\,C\frac{e^{-i\delta_L}}{2}\left[e^{-i(KR-L\pi/2)}-S_L\ e^{i(KR-L\pi/2)}\right]
\eeqn{e36}
where
\beq
S_L&=&e^{2i\delta_L}
\eeqn{e37}
is the \emph{scattering matrix}.

The \Eq{e36} shows that, when we interpret the asymptotic behaviour of these radial scattering wave functions $u_{KL}$ as the sum of an incoming and an outgoing spherical waves, we observe that the outgoing wave is shifted in phase by $2\delta_L$ from the incoming wave.
Intuitively, the former can be seen as describing the particles in the incoming channel, while the latter corresponds to the particles leaving one another after having interacted.
This phaseshift therefore contains all the information available to us on the interaction potential $V$.
Accordingly, following what was said earlier, it can be used to compute the elastic-scattering amplitude, and hence the cross section.

To obtain the scattering amplitude from the phaseshifts, we have to compute the coefficients $c_L$ of the partial-wave expansion \eq{e30}.
This can be achieved formally by comparing the asymptotic behaviour of that expression to that of the stationary scattering wave function \eq{e20} taking \Eq{e36} into account.
Using the partial-wave expansion of plane waves
\beq
e^{iKZ}&\flim{R}{\infty}&\sum_{L =0}^\infty(2L+1)i^LP_L(\cos\theta)\frac{i}{2KR}\left[e^{-i(KR-L\pi/2)}-e^{i(KR-L\pi/2)}\right]
\eeqn{e38}
we obtain
\beq
c_L=\sqrt{4\pi}\sqrt{2L+1}i^Le^{i\delta_L}
\eeqn{e39}
and finally
\beq
f_K(\theta)=\frac{1}{2iK}\sum_{L=0}^\infty(2L+1)(S_L-1) P_L(\cos\theta).
\eeqn{e40}

Following \Eq{e29}, the differential elastic-scattering cross section therefore reads
\beq
\frac{d\sigma}{d\Omega} &=&\left|\frac{1}{2K}\sum_{L=0}^\infty(2L+1)(e^{2i\delta_L}-1) P_L(\cos\theta)\right|^2.
\eeqn{e41}
\noindent After integration over $\Omega$ the total scattering cross section reads
\beq
\sigma=\frac{4\pi}{K^2}\sum_{L=0}^\infty(2L+1)\sin^2\delta_L.
\eeqn{e42a}

\begin{figure}[t]
\center
\includegraphics[height=6cm]{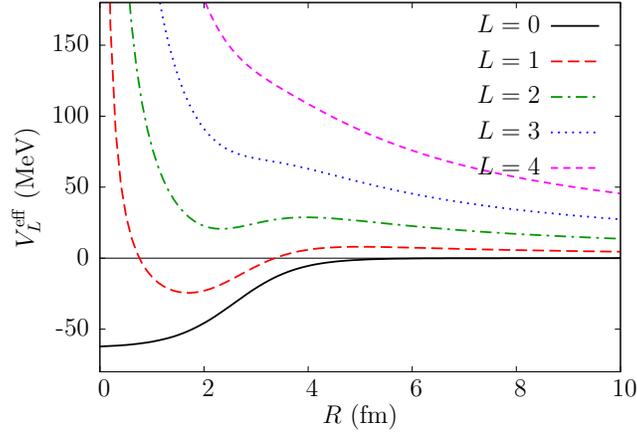}
\caption{\label{f2} Effective potential $V^{\rm eff}$ \eq{e42} plotted for $L=0$ (i.e. $V$), 1, 2, 3 and 4.
We observe that the influence of the interaction potential $V$ on the projectile-target motion decreases as $L$ increases .}
\end{figure}

These expressions show that all the partial waves contribute to the total cross section, however with variable importance depending on the value of the phaseshift $\delta_L$.
To understand this point, let us have a look at the effective potential $V^{\rm eff}_L$, which is the sum of the actual potential $V$ with the centrifugal term $\frac{\hbar^2}{2\mu}\frac{L(L+1)}{R^2}$.
\beq
V^{\rm eff}_L(R)=V(R)+\frac{\hbar^2}{2\mu}\frac{L(L+1)}{R^2}.
\eeqn{e42}
This effective potential is displayed in \Fig{f2} for a typical nucleus-nucleus interaction considering different values of the orbital angular momentum for the projectile-target relative motion $L$.
The purely repulsive centrifugal term combines with the (mostly) attractive potential $V$ to form a \emph{centrifugal barrier} at intermediate distance, where the potential $V$ becomes negligible (here at $R\sim4$~fm).
In a nucleus-nucleus collision as pictured here, this would correspond roughly to the distance at which the surfaces of both nuclei touch one another.
As $L$ increases, the centrifugal barrier increases in height and eventually overcomes the pulling effect of the actual interaction.
Therefore, above a certain value of $L$, the centrifugal term will prevent the colliding particles to come close to one another and therefore to interact with each other.
In that case, the effect of $V$ on the radial wave function $u_{KL}$ will diminish significantly, the phaseshift $\delta_L$ will become very small and hence the contribution of these partial waves to the cross sections \eq{e41} and \eq{e42a} can be neglected.
This is the mechanism that limits the \emph{a priori} infinite sum in the expansion \eq{e30}.

\Fig{f2} also illustrates that the repulsive effect of the centrifugal term will be larger when the $a$-$b$ relative kinetic energy $E$ is lower.
When $E$ gets smaller, the centrifugal barrier gets wider and its top gets higher relative to that incoming energy.
The maximum number of partial waves to include in the expansion hence gets smaller at lower energy.
Eventually, at very low energy, e.g. for reactions of astrophysical interest or for fissions induced by thermal neutrons in conventional nuclear reactors, only one partial wave will matter, the one for which there is no centrifugal barrier, i.e. $L=0$.

\begin{figure}[t]
\center
\includegraphics[trim=1.6cm 13.5cm 1.cm 2cm, clip=true,width=\linewidth]{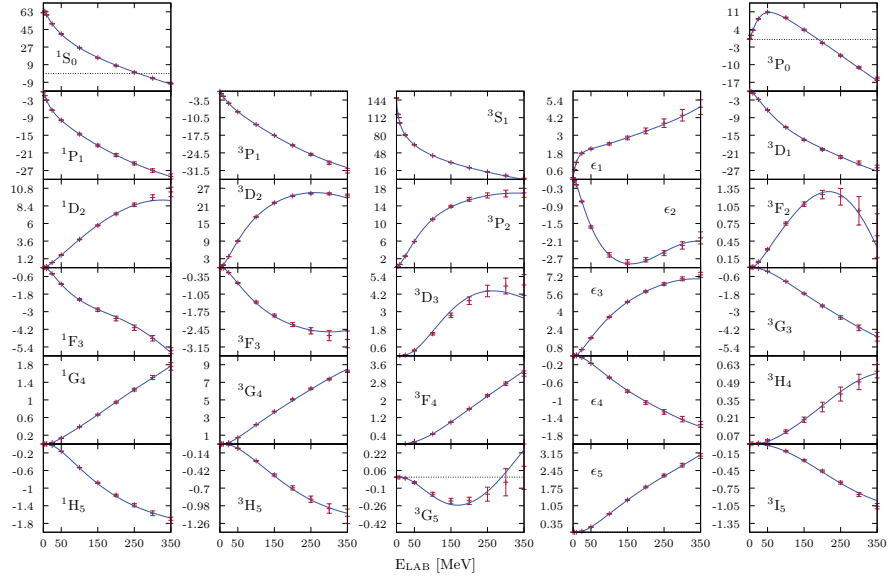}
\caption{\label{f3} Proton-neutron phase shift in various partial waves.
The experimental data (red crosses) are compared with the theoretical prediction of Navarro-P{\'e}rez,  Amaro and Ruiz-Arriola \cite{NAR13}.
Reprinted figure with permission from \Ref{NAR13} Copyright (2013) by Elsevier.}
\end{figure}

To illustrate this effect, I reproduce in \Fig{f3} the proton-neutron phase shift expressed in degrees in different partial waves; this figure is extracted from \Ref{NAR13}.
Each partial wave is identified by the notation $^{2S+1}L_J$, where $S=0$ or 1 is the total spin of the two nucleons, $L$ is their relative orbital angular momentum denoted by a letter ($S\Leftrightarrow L=0$, $P\Leftrightarrow L=1$, $D\Leftrightarrow L=2$, $F\Leftrightarrow L=3$, $G\Leftrightarrow L=4$,\ldots) and $J$ is the total angular momentum obtained from the coupling of $S$ and $L$.

The actual value of these phaseshifts are not important to us.
What matters in this example is to note that when $L$ increases, the range of variation in the phaseshift decreases, going from $180^\circ$ in the $^3S_1$ partial wave down to less that $2^\circ$ in the $H$ and $I$ ones.
This illustrates that the influence of the different partial waves generally decreases as $L$ increases and that this is particularly true at low energy.
At very low energy, the $G$, $H$ and $I$ phaseshifts are indeed negligibly small and can therefore be ignored in the computation of the cross sections.

Before concluding on this partial-wave expansion, let us comment on the analysis performed in \Ref{NAR13}.
As mentioned in \Sec{types}, the study of the collision between two particles can provide information on the interaction between them.
This is exactly what Navarro-P{\'e}rez,  Amaro and Ruiz-Arriola have done here: use the experimental information on the phaseshift in different partial waves in nucleon-nucleon elastic scattering to build a potential that will reliably describe the nucleon-nucleon interaction.
All accurate nucleon-nucleon potentials have been constrained in this way, from the phenomenological ones, like Argonne V18 \cite{AV18} or CD-Bonn \cite{CDBonn} to the more recent nucleon-nucleon interactions derived in chiral effect field theory \cite{EHM09}.

In conclusion, we should therefore see the partial-wave expansion as a low-energy method.
When the incoming energy increases, the number of partial waves to include in the expansion becomes large, and it becomes sensible to consider approximations, like the Born Approximation or the eikonal model of reactions, which have been developed for high-energy reactions (see \Ref{BD04} and \Sec{eikonal}).

\subsubsection{\label{resonances}Resonances}
In the previous section, we have seen that due to the growing influence of the centrifugal term with $L$, the contributions of the partial waves tend to decrease with the orbital angular momentum.
However, there are cases in which one partial wave can have an unexpected dominant contribution over the other ones within a short energy range.
This can happen if the phaseshift in that partial wave $\delta_L$ goes quickly from somewhere close to zero to $\pi/2$  and then up to $\pi$.
In that case, following \Eq{e42a}, we see that the corresponding contribution will go from a small contribution (when $\delta_L$ is small) to its maximum contribution (when $\delta_L=\pi/2$) and then back to something small again (when $\delta_L\sim\pi$).
This behaviour, which corresponds to a significant variation of a cross section within a short energy range, is called a \emph{resonance}.

Because they correspond to a well defined orbital angular momentum $L$ and other quantum numbers (total angular momentum $J$, parity etc.), they can be assimilated to structures in the continuum similar to bound states.
In a sense, they can be seen as a structure composed of the two colliding nuclei sticking to, or orbiting around, one another for a while before separating.

\begin{figure}[t]
\center
\includegraphics[trim=3.8cm 15.4cm 4.4cm 4cm, clip=true, width=5.5cm]{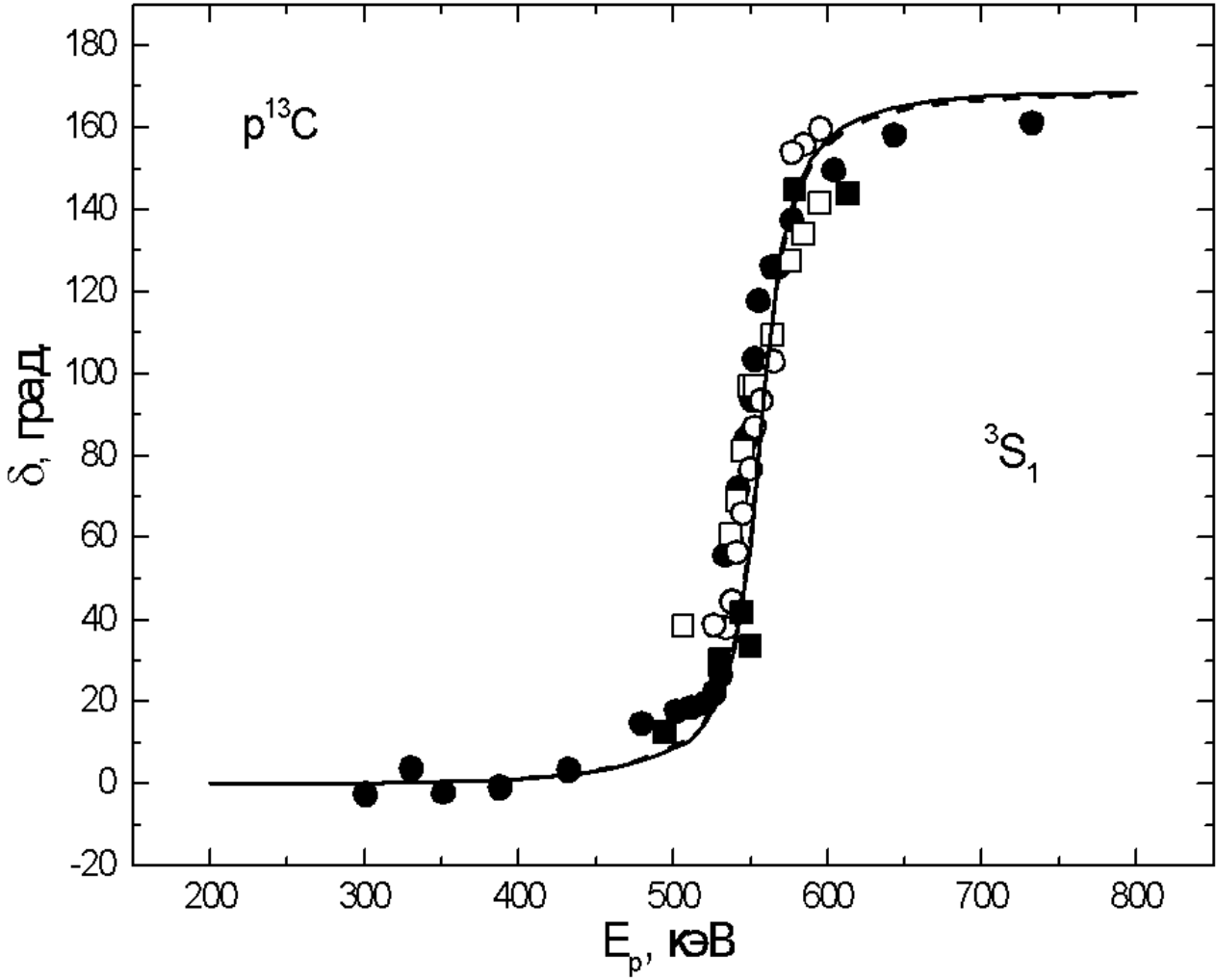}
\hfill
\includegraphics[trim=4.2cm 15.7cm 4.3cm 4cm, clip=true, width=5.5cm]{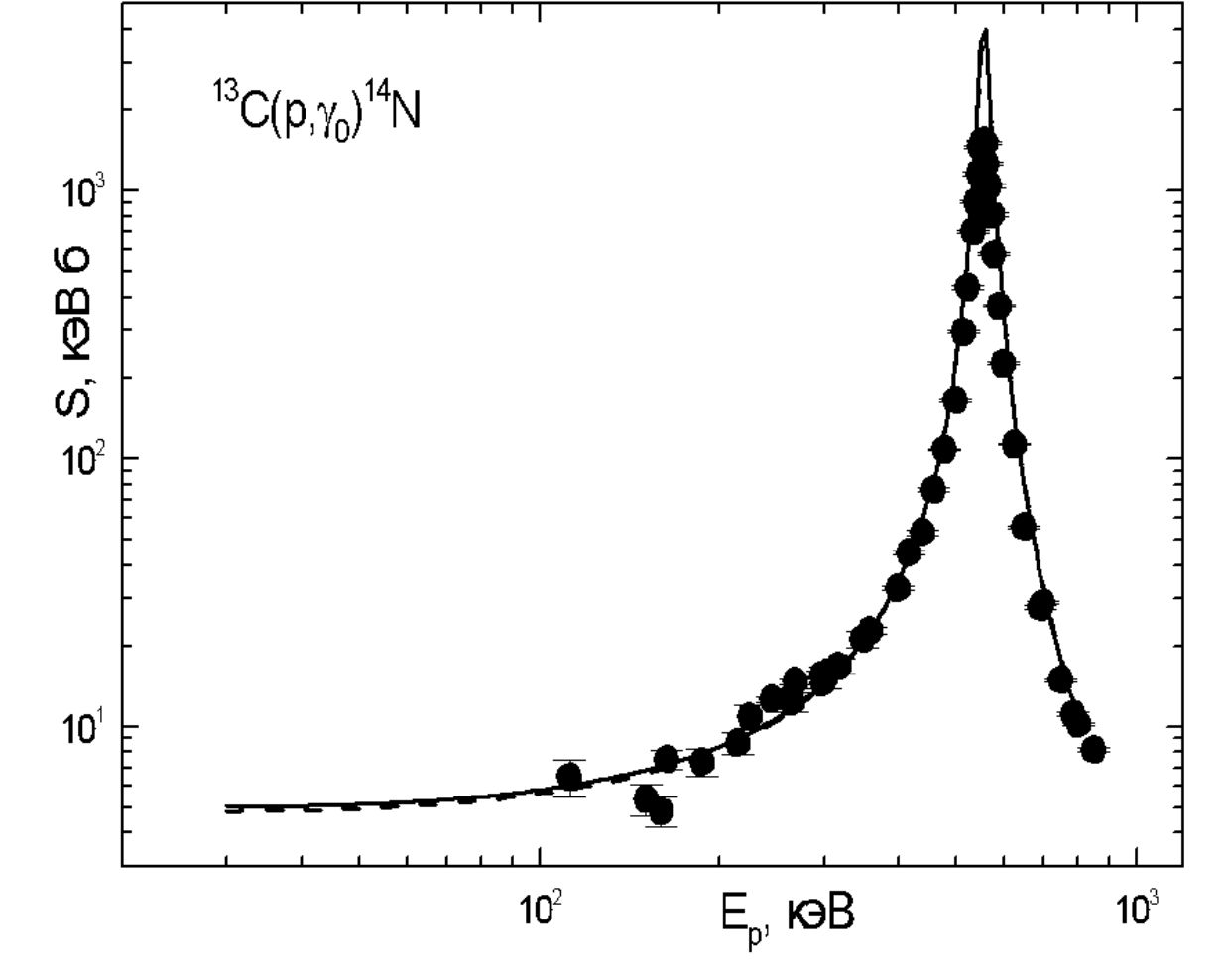}
\caption{\label{f4} Example of a resonance.
Left, the $^3 S_1$ phaseshift in the p-$^{13}$C elastic scattering (in degrees) as a function of the proton energy $E_p$ (in keV). It varies quickly from about $0^\circ$ to nearly $180^\circ$ around $E=550$~keV. Right, this behaviour explains the sudden increase in the cross section for the radiative capture $^{13}$C(p,$\gamma$)$^{14}$N \cite{Dub12}.
Reprinted figure with permission from \Ref{Dub12} Copyright (2012) by Springer Nature.}
\end{figure}

To illustrate this point, \Fig{f4}~(left) reproduces the $^3 S_1$ phaseshift for the elastic scattering of a proton off $^{13}$C as a function of the proton energy $E_{\rm p}$.
We observe a sharp increase in that phaseshift at an energy of about $E_{\rm p}=550$~keV.
The presence of that resonant state eases the capture of the proton by $^{13}$C to form $^{14}$N (a photon $\gamma$ is then emitted to conserve the total energy and momentum of the system).
This is shown in the right panel of \Fig{f4}, where the cross section for that reaction exhibits a sudden and very significant increase at that energy.
This reaction p$+^{13}$C$\rightarrow^{14}$N$+\gamma$, also noted $^{13}$C(p,$\gamma$)$^{14}$N in short, is called a \emph{radiative capture}.

\subsection{\label{OM}Optical Model}
So far, we have focussed only on the elastic-scattering reaction, i.e. when the colliding particles come out of the collision in their initial state, merely scattering each other without change neither in their internal structure nor in their internal energy.
However, as seen in \Sec{types}, other channels can be open.
A cross section can be equally defined for these other channels.
For example, a differential cross section for the transfer $a+b\rightarrow d+e$ can be defined as
\beq
\frac{d\sigma}{d\Omega}(a+b\rightarrow d+e)=\lim_{R\rightarrow\infty}\frac{R^2 J_{d+e}}{J_i},
\eeqn{e43}
where $J_{d+e}$ corresponds to the probability current for the outgoing channel $d+e$, while $J_i$ is the probability flux for the incoming channel \eq{e22}.

We can also define a cross section for various channels.
In particular, the \emph{reaction cross section} ($\sigma_r$) corresponds to the sum of the cross sections for all the channels, but the elastic one:
\beq
\sigma_r=\sum_{{\rm channel}\setminus a+b}\sigma(a+b\rightarrow \mbox{channel}).
\eeqn{e44}
The \emph{interaction cross section} ($\sigma_I$) corresponds to all channels, but the elastic and inelastic scatterings:
\beq
\sigma_I=\sum_{{\rm channel}\setminus (a+b)\cup(a+b^*)\cup(a^*+b)\cup(a^*+b^*)}\sigma(a+b\rightarrow \mbox{channel}).
\eeqn{e45}
This cross section corresponds therefore to all the channels in which the projectile or the target---or both---change their internal structure.
This includes  fusion, transfer, breakup, fragmentation, spallation,\ldots

The existence of other open channels can affect the elastic-scattering process.
In particular, since probability flux appears in these non-elastic channels, some probability flux has to be removed from the elastic-scattering one.
The \emph{optical model} enables us to account for these other channels phenomenologically.

When the interaction between the two particles $a$ and $b$ is described by a real potential, the Hamiltonian $H$ \eq{e18} is unitary and hence the total probability flux $J$ is conserved.
Since that Hamiltonian accounts only for the elastic channel, that probability flux stays in that very channel, even if it shifts from the incoming plane wave to an outgoing spherical wave during the collision [see \Eq{e20}].
This conservation is expressed as the continuity equation, which, for a stationary state, reads
\beq
\ve{\nabla}\ve{J}(\ve{R})=0
\eeqn{e46}
To simulate absorption of the probability flux towards other channels, it has been suggested to use a complex---or \emph{optical}---potential
\beq
U_{\rm opt}(R)=V(R)+i\ W(R).
\eeqn{e47}
Usually, but not always, optical potentials are referred to by the symbol $U$, instead of $V$, which is often kept for real potentials or, as in \Eq{e47}, for the real part of complex potentials.
The imaginary part of these potentials is usually denoted by $W$.

When the interaction between the colliding particles is complex, the divergence of the probability flux is no longer nil.
Accounting for the \Sch equation \eq{e19} written with the optical potential $U_{\rm opt}$ instead of the purely real one used so far, we obtain
\beq
\nabla \ve{J}(\ve{R})&=& \nabla \frac{1}{\mu}\Re\{\Psi^*(\ve{R})\,\ve{P}\,\Psi(\ve{R})\}\nonumber\\
&=& -i\frac{\hbar}{2\mu}\nabla [\Psi^*(\ve{R})\,\nabla\Psi(\ve{R})-\Psi(\ve{R})\,\nabla\Psi^*(\ve{R})]\nonumber\\
&=&i\frac{\hbar}{2\mu}[\Psi^*(\ve{R})\,\frac{2\mu}{\hbar^2}U_{\rm opt}(R)\Psi(\ve{R})-\Psi(\ve{R})\,\frac{2\mu}{\hbar^2}U_{\rm opt}^*(R)\Psi^*(\ve{R})]\nonumber\\
&=&\frac{2}{\hbar}W(R)\left|\Psi(\ve{R})\right|^2.
\eeqn{e48}
For that divergence to model \emph{absorption} from the elastic channel, it has to be negative.
Therefore, the imaginary part of optical potentials has to be negative (or nil): $W(R)\le0 \ \forall\ve{R}$.

To understand how this affects the wave function, let us have a look at the partial-wave expansion explained in the previous section.
From \Eq{e33}, we can see that using a complex optical potential in the calculation will lead to a complex phaseshift $\delta_L$.
It can be shown that since the imaginary part of the potential is negative, the imaginary part of the phaseshift is positive [$\Im(\delta_L)\ge0$].
The scattering matrix \eq{e37} then reads
\beq
S_L&=&\eta_L\ e^{2i\Re(\delta_L)},
\eeqn{e49}
where $\eta_L=e^{-2\Im(\delta_L)}\le1$.
Accordingly, the asymptotic behaviour of the radial wave function $u_{KL}$ \eq{e36} exhibits an outgoing spherical wave with a reduced amplitude compared to the incoming one
\beq
u_{KL}&\flim{R}{\infty}&\propto\left[e^{-i(KR-L\pi/2)}-\eta_L\ e^{2i\Re(\delta_L)}\ e^{i(KR-L\pi/2)}\right].
\eeqn{e50}
This clearly corresponds to a loss of flux from the entrance channel, simulating the presence of other open channels that can be populated during the collision.
The name \emph{optical model} comes from optics, where a complex index of refraction can be used to simulate the absorption of light by the medium.

Within this model, one can compute an \emph{absorption cross section} $\sigma_a$ that corresponds to all non-elastic channels.
It can be computed as the elastic-scattering cross section in \Eq{e28} or the transfer cross section in \Eq{e43}.
Alternatively, we can also integrate  the loss of flux \eq{e48} over the whole space:
\beq
\sigma_a&=&\frac{-\int \nabla\ve{J}\ d\ve{R}}{J_i} \label{e51}\\ 
 &=&\frac{-\lim_{R\rightarrow\infty}\oint \ve{J}\cdot\ve{\hat R}\ R^2d\Omega}{J_i}\nonumber\\
 &=&\frac{\pi}{K^2}\sum_{L=0}^\infty(2L+1)\left(1-\eta_L^2\right).
\eeqn{e52}
By analogy to \Eq{e50}, let us note that since $\eta^2_L$ measures the flux of probability that remains in the elastic channel, $1-\eta_L^2$ corresponds to what has been absorbed from that channel.
At the limit of a real potential, $\eta_L=1\ \forall L$ and the absorption cross section is nil.
From what has been introduced at the beginning of this section, we see that $\sigma_a$ can be compared to the reaction cross section $\sigma_r$, which can be measured experimentally.

Note that all the developments performed in this section have been made for a short-range potential, i.e. for which $R^2V(R)\flim{R}{\infty}0$.
This is not valid for a Coulomb potential, which is non-negligible in most of the nuclear-physics problems, since nuclei are charged.
Fortunately, although it leads to more complicated calculations, the Coulomb part of the interaction can be treated exactly and results similar to what has been shown here can be performed seeing the nuclear interaction as a short-ranged perturbation to the Coulomb interaction.
I refer the interested readers to more detailed references to study the corresponding developments \cite{Tay72,NT09,BD04}.

\section{\label{bu}Breakup Models}

\subsection{\label{halo}An Introduction to Halo Nuclei}

Stable nuclei exhibit a compact structure with a density (for both matter and charge) roughly constant from the centre of the nucleus until its surface.
Beyond that point, the density drops quickly with a decay similar throughout the whole nuclear chart.
This structure leads to a \emph{liquid-drop} model of the nucleus \cite{Kra87}, in which the nucleons are seen tightly packed, interacting mostly to their closest neighbours and forming a nuclear droplet of constant density.
Their volume being directly proportional to the number $A$ of nucleons, the radius of these drops hence vary linearly with $A^{1/3}$.

In the mid-80s, when the first Radioactive Ion Beams became available, Tanihata and his collaborators have undertaken to test if this property remains true away from stability.
Because radioactive nuclei are, by definition, unstable, it is not possible to measure their radii with usual techniques, like electron scattering.
Therefore, to estimate the size of these nuclei, they have measured their interaction cross section $\sigma_I$ (see \Sec{OM}) on various targets at high energy \cite{Tan85b,Tan85r}.
In a very geometric model, where the projectile $P$ and the target $T$ are seen as hard spheres, they will \emph{interact} in the sense of $\sigma_I$ when they touch each other.
The cross section for this process thus reads
\beq
\sigma_I(P,T)=\pi[R_I(P)+R_I(T)]^2,
\eeqn{e53}
where $R_I$ is the \emph{interaction radius}.
In a first approximation, this radius can be used to estimate the size of the nucleus.

\begin{figure}[b]
\center
\includegraphics[width=8cm]{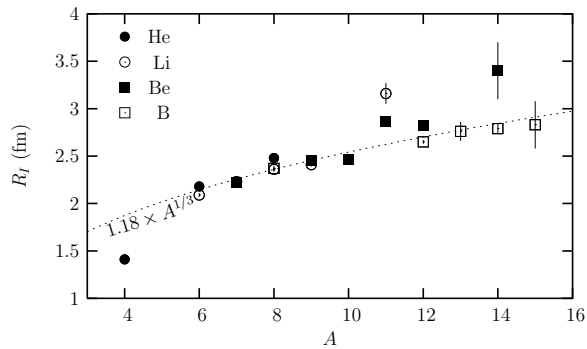}
\caption{\label{f5} Interaction radius for isotopes of He, Li, Be and B. Data are from \Ref{Tan96}.}
\end{figure}

\Fig{f5} shows the interaction radii obtained by Tanihata \etal\ for various isotopes of He (full circles), Li (open circles), Be (full squares) and B (open squares) \cite{Tan96}.
In addition to the data, the $A^{1/3}$ usual behaviour is plotted as a dotted line.
We can see that most of the nuclei studied here exhibit more or less the radius predicted by this empirical law.
However, a few of them, like $^{11}$Li, $^{11}$Be or $^{14}$Be stick out of this trend and seem larger than expected.
One possibility to explain this unusual feature could be a significant collective deformation.
However, an exotic structure is actually the reason for these variations from a well established property of stable nuclei.

Subsequent studies have shown that the valence neutrons play a significant role in the sudden increase of the interaction cross section for these nuclei.
For example, it has been observed that the difference between the interaction cross section of one of these exotic nuclei and that of the isotope one or two neutron down the isotopic line roughly equals the one- or two-neutron removal cross section for that very nucleus \cite{Tan96}.
This suggests that most of the increase in the interaction cross section is due to the additional valence neutrons.

Other groups have then measured one-neutron knockout (KO) on these nuclei.
In that reaction, one neutron is removed from the projectile during a high-energy collision on a light target, like C or Be (see, e.g., \Ref{Sau00}).
In these measurements, only the $A-1$ nucleus produced by the knockout is measured, not the kicked out neutron.
For the nuclei that exhibit an unusually large interaction radius, experimentalists have observed narrow parallel-momentum distributions of the remaining $A-1$ nucleus.
Since the reaction takes place at high energy, one can assume that the measured distribution reveals the momentum distribution the remnant cluster had within the initial nucleus.
That distribution being narrow, it means, from the Heisenberg uncertainty principle, that the spatial neutron-core distribution must be extended.

These different results---large interaction cross section, special role of the valence neutrons, narrow momentum distribution of the $A-1$ nucleus in KO,\ldots---have led to the notion of \emph{halo nuclei} \cite{HJ87}.
Halo nuclei are light, neutron-rich nuclei that are unusually large compared to their isobars, viz. their matter radius does not follow the empirical $A^{1/3}$ rule.
This large size is qualitatively understood as resulting from their small one- or two-neutron separation energies ($S_{\rm n}$ or $S_{2\rm n}$).
Thanks to this loose binding, the valence nucleons can tunnel far into the classically forbidden region, i.e. far away from the range of the nuclear interaction.
They hence form a sort of diffuse \emph{halo} around the core of the nucleus, which exhibits a usual nuclear structure, being tightly bound and compact.
Quantum-mechanically, this translates by a long-range tail in the wave function that describes their relative motion to the other nucleons.

Two types of halo nuclei have been observed so far: the one-neutron halo nuclei, and those with two neutrons in their halo.
$^{11}$Be and $^{15}$C exhibit a one neutron halo; they are thus seen as a $^{10}$Be, respectively $^{14}$C, core to which one neutron is loosely bound.
$^{6}$He and $^{11}$Li are typical examples of two-neutron halo nuclei; they exhibit a clear three-body structure with an $\alpha$, respectively $^9$Li, core surrounded by two valence neutrons.

\begin{figure}[t]
\center
\includegraphics[width=5.5cm]{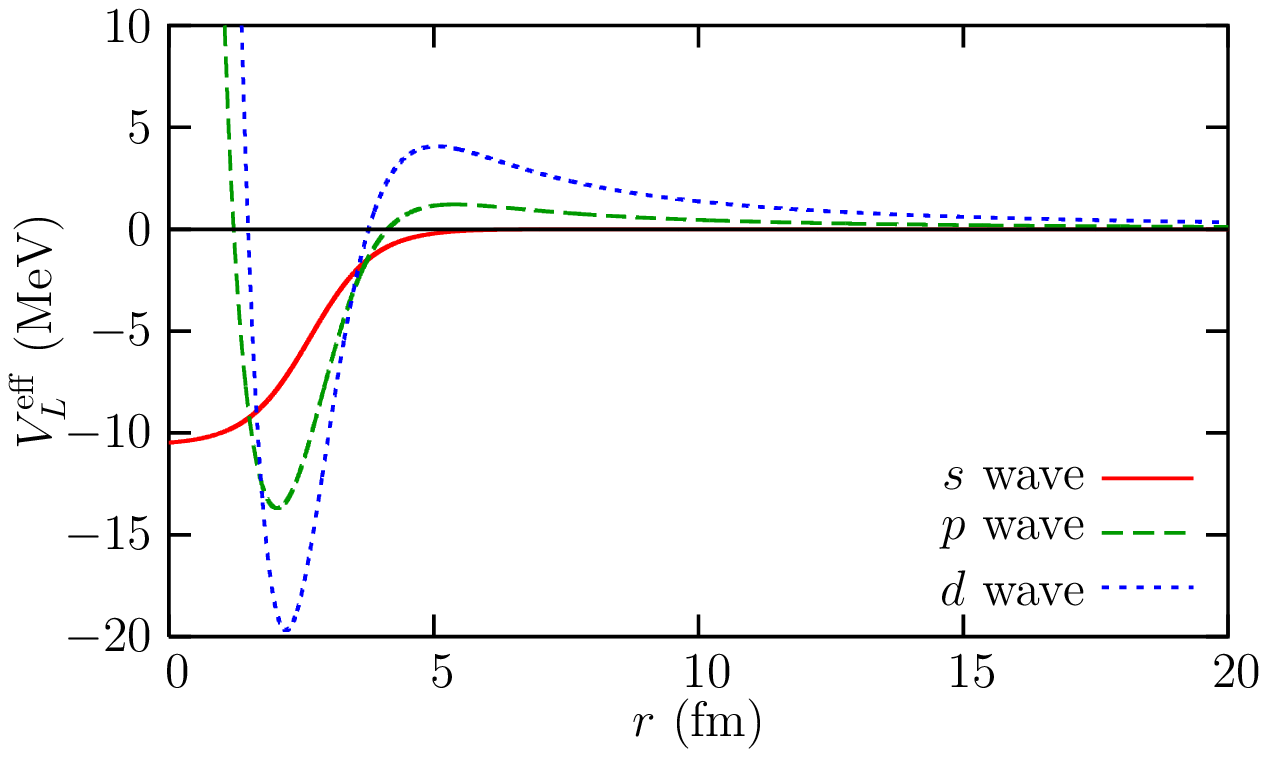}
\includegraphics[width=5.5cm]{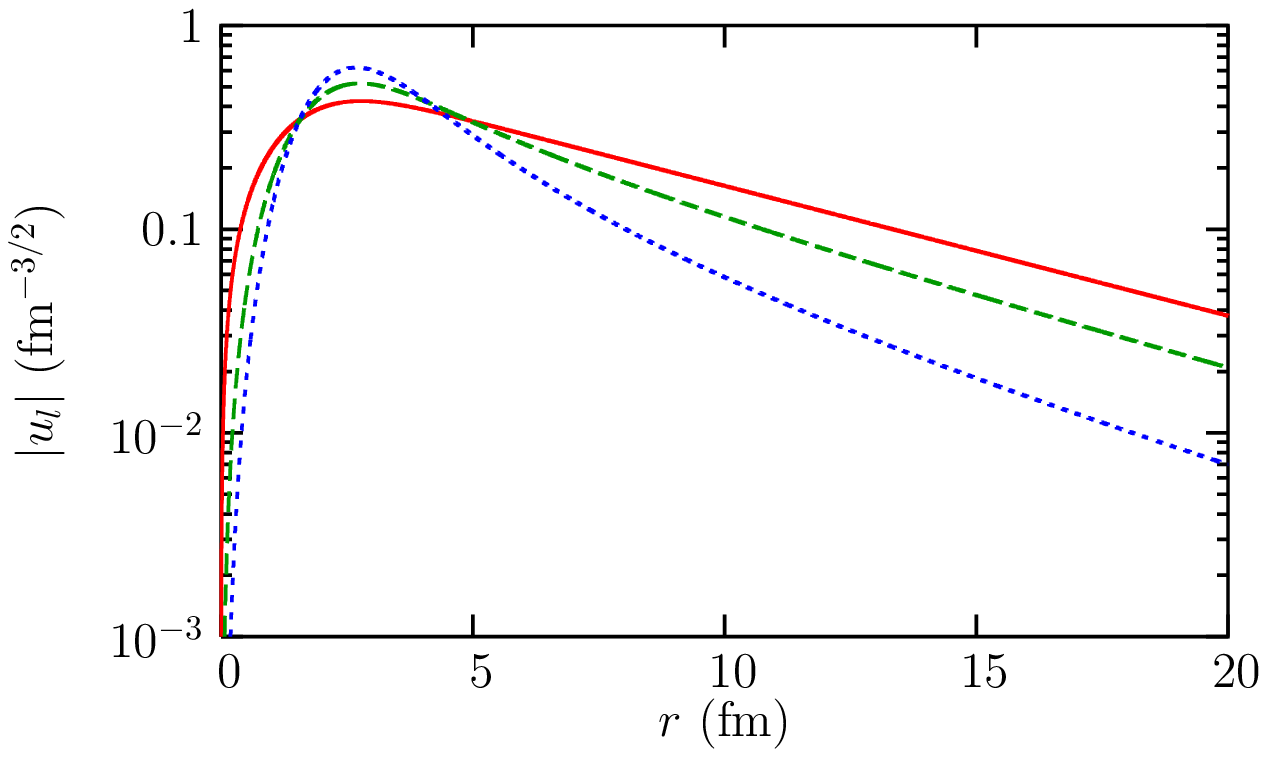}
\caption{\label{f6} Left: Effective potential $V_L^{\rm eff}$ \eq{e42} in $S$, $P$ and $D$ waves fitted to host a state bound by 0.5~MeV.
Right: Reduced radial wave functions of the corresponding bound states.}
\end{figure}

In order for a halo to develop in a nucleus, not only does $S_{\rm n}$ or $S_{2\rm n}$ need to be small, but also nothing may hinders the extension of the wave function to large distances.
Therefore halos are observed mostly when the valence neutrons sit in an orbital corresponding to a low orbital angular momentum $L$, i.e. usually in an $S$ or $P$ wave.
This is illustrated in \Fig{f6}, where effective potentials $V_L^{\rm eff}$ \eq{e42} in the $S$, $P$ and $D$ waves are plotted as a function of the radial distance $R$ (left).
Each of them is fitted to host a state bound by $0.5$~MeV, which is the one-neutron separation energy of $^{11}$Be, the archetypical one-neutron halo nucleus.
The corresponding radial wave functions are depicted in \Fig{f6} (right).
We observe that a higher orbital angular momentum will lead to a higher centrifugal barrier, which will force the wave function inside the nucleus, hence reducing the probability that the valence neutron can be found far from the other nucleons.

For this reason, although not impossible, the development of a halo in proton-rich nuclei is less probable.
In that case, in addition to the centrifugal barrier, the valence nucleon always feels a Coulomb barrier, that also reduces the density of probability at large distance.
The ground state of $^{8}$B and the first excited state of $^{17}$F are often seen as candidates for one-proton halo nuclei.

Note that in addition to exhibiting a halo, two-neutron halo nuclei, like $^{6}$He or $^{11}$Li, are also \emph{Borromean nuclei}.
This means that although the three-body system composed of the core plus two neutrons is bound, none of the two-body subsystems is: although $^6$He exists, neither the dineutron $^2$n nor $^5$He exist \cite{Zhu93}.
This name was coined by Zhukov \etal\ after the Borromean rings, which are three rings entangled in such a way that when one of the rings is broken, the other two get loose.
These rings appear on the coat of arms of the Borromeans, a noble family from Northern Italy, hence their name.

As mentioned above, halo nuclei are usually located close to the neutron dripline, i.e. on the edge of the  valley of stability.
These nuclei are therefore (very) short-lived and cannot be studied through usual spectroscopic methods, like electron or proton elastic scattering.
To probe their internal structure, we must then rely on indirect techniques, like reactions.
To infer reliable nuclear-structure information about the projectile from the experimental data, a precise model of the reaction coupled to a realistic description of the projectile must be used.
In the next section, I describe different models that exist for \emph{breakup reactions}.

\subsection{Breakup Reaction}

\subsubsection{Introduction}
In breakup reactions the projectile dissociates into its more elementary constituents during its collision with a target (see also \Sec{types}).
Such process takes place because the constituents of the projectile interact differently with the target, leading to a tidal force that can be sufficient to break the nucleus apart.
This reaction hence reveals the internal structure of the projectile.
It is particularly well suited to study the cluster structure of loosely bound nuclei, like halo nuclei.
When performed on heavy targets, i.e. when they are dominated by the Coulomb interaction, breakup reactions can also be used to infer reaction rates of astrophysical interest \cite{BBR86,BR96}.

In what follows, we will focus on what is called the \emph{elastic} or \emph{diffractive} breakup, i.e. the reaction in which all the clusters are measured in coincidence.
In that case we talk of \emph{exclusive} measurements, in opposition to the \emph{inclusive} measurements, in which only some of the outgoing particles are detected, like in knockout.

\subsubsection{\label{framework}Theoretical Framework}
Since breakup leads to the dissociation of the projectile into two or more parts, the most basic descriptions of that reaction must include these parts as degrees of freedom.
The simple model presented in \Sec{scatt}, in which the internal structure of the colliding nuclei is neglected, is therefore not sufficient to describe the dissociation of the projectile.
In this introduction to breakup modelling, we will assume the simplest case of a two-cluster projectile: a core $c$ to which a fragment $f$, e.g. a valence nucleon, is loosely bound.
This corresponds to reactions involving one-nucleon halo nuclei, like $^{11}$Be or $^{8}$B.

Such a projectile is described phenomenologically by the one-body Hamiltonian
\beq
H_0=T_r+V_{cf}(\ve{r}),
\eeqn{e54}
where $\ve{r}$ is the coordinate of the fragment relative to the core and $T_r=-\hbar^2\Delta_r/2\mu_{cf}$ is the operator corresponding to the $c$-$f$ kinetic energy, with $\mu_{cf}=m_cm_f/m_P$ the $c$-$f$ reduced mass, where $m_c$ and $m_f$ are the masses of the core and the fragment, respectively, and $m_P=m_c+m_f$ is the mass of the projectile.
In \Eq{e54}, $V_{cf}$ is a phenomenological potential that simulates the interaction that binds the fragment to the core.

In this simple picture, the $c$-$f$ relative motion is described by the eigenstates of the Hamiltonian $H_0$ \eq{e54}.
Since the potential $V_{cf}$ is usually chosen as central, the $c$-$f$ orbital angular momentum $l$ and its projection $m$ are good quantum numbers (the spin is ignored here for clarity; although a bit tedious, the extension of the developments of this section to the case of particles with spin is not difficult).
In partial waves $lm$ these states are thus solution of
\beq
H_0\ \phi_{lm}(\ve{r}) = E\ \phi_{lm}(\ve{r}),
\eeqn{e55}
where $E$ is the $c$-$f$ relative energy.
The threshold $E=0$ corresponds to the fragment separation from the core.
Negative energies ($E<0$) correspond to states in which the fragment is bound to the core, while positive-energy states ($E>0$) describe the projectile broken up into its core and fragment.
The former states are discrete; we add the number of nodes in the radial wave function $n$ to $l$ and $m$ to enumerate them: $\phi_{nlm}$.
Since the latter states describe the fragment separated from the core, they correspond to a \emph{continuum} of energies, possibly including resonances (see \Sec{resonances}).
We add the wave number $k$ ($E=\hbar^2k^2/2\mu_{cf}$) to $l$ and $m$ to distinguish them and remind that they belong to the continuum: $\phi_{klm}$.

The $c$-$f$ potential $V_{cf}$ usually exhibits a Woods-Saxon form sometimes with a spin-orbit coupling term.
Its parameters (mostly its depth) are adjusted to reproduce the know low-energy spectrum of the nucleus, i.e. the binding energy of the fragment to the core, the spin and parity of the ground state and maybe some of the excited states as well, including sometimes resonances.

As an example, let us mention the case of $^{11}$Be, which I will use later to illustrate different models of reactions.
Being the archetypical one-neutron halo nucleus, $^{11}$Be is usually described as a $^{10}$Be core to which a neutron is bound by a mere $S_{\rm n}(^{11}{\rm Be})=501.64\pm0.25$~keV \cite{NNDC}.
Its ground state has spin and parity $\half^+$.
In addition, it also has an excited $\half^-$ bound state with $S_{\rm n}(^{11}{\rm Be}^*)=181.60\pm0.35$~keV \cite{NNDC,KKP12}.
Above the one-neutron separation threshold, at a $c$-$f$ energy $E=1.281$~MeV~$\pm4$~keV \cite{KKP12}, $^{11}$Be exhibits a $\fial^+$ resonance.
Assuming that the $^{10}$Be core is in its $0^+$ ground state, the $\half^+$ state of $^{11}$Be is described as a neutron bound to the core in the $1s_{1/2}$ orbit, where we have added to the number of radial nodes $n=1$ and the orbital angular momentum $l=0$, the total angular momentum $j$ obtained from the coupling of $l$ to the spin $s=\half$ of the valence neutron, hence $j=\half$ here.
In that model, the $\half^-$ excited state is seen as a $0p_{1/2}$ neutron bound to $^{10}$Be($0^+$) and the $\fial^+$ resonance is usually reproduced in the $d_{5/2}$ orbital.
In this way the energy relative to the one-neutron separation threshold, the total angular momenta and the parity of these states agree with the known low-energy spectrum of $^{11}$Be.
Note that this enables us to constrain the potential $V_{cf}$ in partial waves $s_{1/2}$, $p_{1/2}$, and $d_{5/2}$, but not in the $p_{3/2}$ or higher.
This has significant implications in the analysis of actual breakup data (see \Sec{structure}).

\begin{figure}[t]
\sidecaption[t]
\includegraphics[width=4cm]{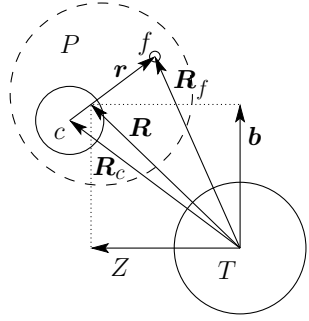}\ \ \ \ \ \ \ \ \ \ \ \ \ \ \ \ \ \ \ \ \ \ \ \ \ \ \ \ \ \ \ \ \
\caption{\label{f7} Jacobi set of coordinates $(\ve{r},\ve{R})$ used to describe the breakup on a target $T$ of a projectile $P$ composed of two internal clusters, a core $c$ and a fragment $f$.
The beam axis is usually chosen as the $Z$ axis.
The transverse coordinate of $\ve{R}$ is denoted by $\ve{b}$.
The $c$-$T$ ($\ve{R}_c$) and $f$-$T$ ($\ve{R}_f$) relative coordinates are shown for completeness.}
\end{figure}

The third body in the model of the reaction is the target $T$.
It is usually described as a structureless particle, whose interaction with the projectile constituents are simulated by optical potentials $U_{cT}$ and $U_{fT}$ (see \Sec{OM}).
Once the three-body centre of mass motion has been removed, we are left with the Jacobi set of coordinates illustrated in \Fig{f7}.
It is composed of the coordinate of the projectile centre of mass relative to the target $\ve{R}$, in addition to the $c$-$f$ coordinate $\ve{r}$ used in \Eq{e54}.

\newpage

Within this three-body model, describing the $P$-$T$ collision reduces to solving the following \Sch equation
\beq
\left[T_R+H_0+U_{cT}(\ve{R}_c)+U_{fT}(\ve{R}_f)\right]\Psi(\ve{r},\ve{R})=E_T\Psi(\ve{r},\ve{R}),
\eeqn{e56}
where $T_R=-\hbar^2\Delta_R/2\mu$ is the $P$-$T$ kinetic-energy operator, with $\mu$ the $P$-$T$ reduced mass, and $\ve{R}_c$ and $\ve{R}_f$ are the $c$-$T$ and $f$-$T$ relative coordinates, respectively (see \Fig{f7}).
In the center-of-mass rest frame, the total energy is related to the initial $P$-$T$ relative momentum $\hbar K$ and the binding energy $E_{n_0l_0}$ of the projectile ground state $\phi_{n_0l_0m_0}$: $E_T=\hbar^2K^2/2\mu+E_{n_0l_0}$.

The \Eq{e56} has to be solved with the condition that the projectile, initially in its ground state $\phi_{n_0l_0m_0}$, is impinging on the target:
\beq
\Psi^{(m_0)}(\ve{r},\ve{R})\flim{Z}{-\infty}e^{iKZ}\ \phi_{n_0l_0m_0}(\ve{r}).
\eeqn{e57}
It is not possible to solve this three-body problem exactly.
Therefore, various methods have been developed over the years to solve it numerically with more or less sophistication and using different levels of approximation.
We will see three of them in the next sections: the Continuum Discretised Coupled Channel method (CDCC), the time-dependent approach (TD) and the eikonal approximation.
A recent and more complete review of these methods can be found in \Ref{BC12}.

\clearpage

\subsubsection{\label{cdcc}Continuum Discretised Coupled Channel Model (CDCC)}

Since the eigenstates $|\phi_i\rangle$ of the Hamiltonian $H_0$ \eq{e54} form a basis in the vector space of the projectile internal coordinate $\ve{r}$, the idea of the \emph{Continuum Discretised Coupled Channel} method (CDCC) is to expand the three-body wave function $\Psi$ upon that basis
\beq
\Psi(\ve{r},\ve{R})=\sum_i\chi_i(\ve{R})\langle\ve{r}|\phi_i\rangle.
\eeqn{e58}
Introducing this expansion in the \Sch equation \eq{e56} leads to
\beq
\sum_i T_R\chi_i(\ve{R})|\phi_i\rangle+\chi_i(\ve{R})H_0|\phi_i\rangle+\left[U_{cT}(\ve{R}_c)+U_{fT}(\ve{R}_f)\right]\chi_i(\ve{R})|\phi_i\rangle= \sum_iE_T\chi_i(\ve{R})|\phi_i\rangle.\nonumber\\
\eeqn{e59}
Accounting for the fact that $H_0|\phi_i\rangle=E_i|\phi_i\rangle$ and projecting each member of the equality on $\langle\phi_j|$ one gets the equations the functions $\chi_j$ of $\ve{R}$ must satisfy
\beq
T_R\,\chi_j(\ve{R})+\sum_i\langle\phi_j|U_{cT}+U_{fT}|\phi_i\rangle\,\chi_i(\ve{R})=(E_T-E_j)\,\chi_j(\ve{R}).
\eeqn{e60}
This is a set of \emph{coupled equations} in which the coupling terms are the matrix elements of the optical potentials $U_{cT}+U_{fT}$ within the basis of the eigenstates of $H_0$ $|\phi_i\rangle$.
These terms simulate the interaction between the projectile constituents and the target.
Without them nothing would happen and the projectile would pass by the target unscathed, and the $P$-$T$ system would stay in the initial channel $e^{iKZ}\phi_{n_0l_0m_0}$.
As mentioned above, because of these interactions, the system can shift from this initial (elastic) channel to other channels, where the projectile is in other eigenstates of $H_0$.
A stated in \Sec{framework}, these states can correspond to excited bound states of the projectiles or, in the case of breakup, to the $c$-$f$ continuum, i.e. one of the states $\phi_{klm}$.
Because they describe a continuum part of the spectrum, these states are not numerically tractable.
The symbolic expansion \eq{e58} should actually include an integral over the value of the $c$-$f$ relative momentum $\hbar k$.
To circumvent this issue, Rawitscher has suggested to \emph{discretise} this continuum \cite{Raw74}, viz. to approximate the continuum functions $\phi_{klm}$ by a set of discrete $\phi_{ilm}$.
This hence leads to the Continuum Discretised Coupled Channel method or CDCC \cite{Kam86,Aus87}, see \Ref{Yah12} for a recent review.

Various methods have been suggested to discretise the continuum.
The simplest idea is to divide the continuum into small energy intervals, also known as \emph{bins}: $\left[E_i-\Delta E_i/2,E_i+\Delta E_i/2\right]$ and choose to describe each interval by the continuum wave function at the midpoint:
\beq
\phi_{ilm}(\ve{r})=\phi_{k_ilm}(\ve{r}),
\eeqn{e61}
 with $E_i=\hbar^2k_i^2/2\mu_{cf}$.

Unfortunately, these mid-point functions are not square integrable (as seen in \Sec{phaseshift} they oscillate indefinitely when $r\rightarrow\infty$).
It has then been suggested to choose as bin functions the average
\beq
\phi_{ilm}(\ve{r})=\frac{1}{W_{il}}\int_{E_i-\frac{\Delta E_i}{2}}^{E_i+\frac{\Delta E_i}{2}}f_{l}(E)\ \phi_{klm}(\ve{r})\,dE,
\eeqn{e62}
where $f_{l}$ is a weight function that can differ from one partial wave to another and the normalisation factor $W_{il}=\int_{E_i-\frac{\Delta E_i}{2}}^{E_i+\frac{\Delta E_i}{2}}f_{l}(E) dE$.
This method produces square-integrable wave functions, which ease the calculation of the coupling terms in \Eq{e60} \cite{Raw74,Kam86,Aus87}.
However, these bin wave functions may extend over a long distance before becoming negligible, especially if the bins are narrow.

A third way to obtain a discretised continuum is to use \emph{pseudo-states}.
These states can be generated in different ways.
A first one is by diagonalising the projectile Hamiltonian $H_0$ within a finite set of square-integrable basis states, e.g. using an $R$-matrix formalism \cite{DBD10}.
The resulting eigenstates of $H_0$ are thus also discrete and square integrable, even for positive energies $E$.
Another way to obtain pseudo-states is to use a Transformed Harmonic Oscillator (THO) basis \cite{THO}.
The idea of this method is to build a mathematical transformation between the harmonic-oscillator states and the eigenstates of $H_0$ \eq{e55}.
Since the eigenstates of the harmonic oscillator are naturally discrete and square integrable, the transformation produces the desired discretised continuum.

The pseudo-states produced by these different methods usually vanish faster at large $r$ than the radial wave function obtained by averaging \eq{e62}, which is useful in the calculation of the coupling matrix elements of \Eq{e60}.
However, unlike in the binning technique, the eigenenergies $E_i$ cannot be chosen at will because they are the direct outcome of the diagonalisation of the Hamiltonian $H_0$, and hence depend on the choice of the basis states considered, or the way the THO is built.


Except for this discretisation of the continuum, the CDCC method does not make any other approximation and solves the three-body \Sch equation \eq{e56} ``exactly'', viz. numerically.
It is therefore fully quantal and makes no approximation on the $P$-$T$ relative motion.
It can thus be used at all energies.
However, it may be quite computationally expensive, especially at high energies.
This is the reason why other models based on different approximations of the $P$-$T$ relative motion have been suggested (see Secs.~\ref{TD} and \ref{eikonal}).

Various codes have been written to solve the CDCC coupled equations.
The code {\sc fresco} written by Thomson is the best known \cite{fresco}; it can be freely downloaded from \url{www.fresco.org.uk}.

\begin{figure}[t]
\center
\includegraphics[width=8cm]{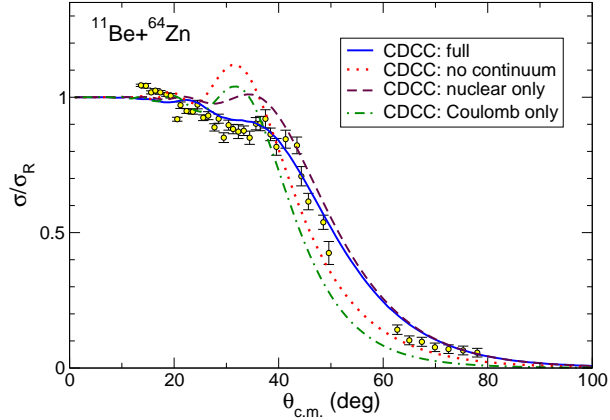}
\caption{\label{f8} Elastic scattering of $^{11}$Be on Zn at $24.5$~MeV measured at ISOLDE (CERN) \cite{Dip10,Dip12}.
The coupling to the $^{10}$Be-n breakup channel included in the ``full'' CDCC calculation must be accounted for in order to properly reproduce the measured cross section.
Reprinted figure with permission from \Ref{Dip12} Copyright (2012) by the
American Physical Society.}
\end{figure}

To illustrate the CDCC method, I display in \Fig{f8} the elastic-scattering cross section of $^{11}$Be on Zn at 24.5~MeV \cite{Dip10,Dip12}.
This measurement is part of a larger effort to measure the elastic scattering of beryllium isotopes $^{9,10,11}$Be on Zn around the $P$-$T$ Coulomb barrier to see if the presence of the one-neutron halo in $^{11}$Be has any effect on the elastic process.
The stable $^{9}$Be and near-to-stability $^{10}$Be exhibit usual elastic-scattering cross sections, which can be easily described with usual optical potentials (see \Sec{OM}).
The halo nucleus $^{11}$Be, however, exhibits a significant suppression of its elastic-scattering cross section around $\theta\sim30^\circ$--$40^\circ$ (see \Fig{f8}), which can only be reproduced if a long-range term is added to the imaginary part of the optical potential \cite{Dip10}.
This suggests that the halo indeed influences the elastic scattering.
To investigate in more details this effect, precise CDCC calculations of this collision have been performed \cite{Dip12}.

To properly reproduce the data, a full calculation, i.e., including the coupling to and within the $^{10}$Be-n continuum needs to be included (solid blue line in \Fig{f8}).
Solely accounting for the extended wave function of the projectile without the couplings to the continuum (red dotted line in \Fig{f8}) does not exhibit the strong suppression of the elastic-scattering cross section at $30^\circ$.
This shows the importance to account for the breakup channel in collision involving loosely bound systems, like halo nuclei, even if the breakup process is not the primarily study of the experiment.

\subsubsection{\label{TD}Time-Dependent Approach}
To reduce the computational complexity of the CDCC framework, various approximations have been suggested.
The first one I present in this short review is the \emph{Time-Dependent} approach (TD).
The main idea is to approximate the $P$-$T$ relative motion by a classical trajectory $\ve{R}(t)$, while maintaining a quantum description of the projectile.
This approach was first developed to describe Coulomb excitation in heavy-ion collisions, in which a nucleus is excited during its collision with a heavy target \cite{AW75}.
The high $Z$ of the target ensures that the process is dominated by the Coulomb interaction.
The reaction can thus be described as resulting from the exchange of virtual photons between the colliding nuclei.
It has naturally be extended to describe the Coulomb breakup of halo nuclei, where the excitation takes place between the initial ground state of the projectile \eq{e57} and its continuum \cite{KYS94,EBB95,TW99,Fal02,CBM03c}.

In this approximation, the projectile follows a classical trajectory $\ve{R}(t)$ along which it feels a time-dependent potential simulating its interaction with the target.
This leads to the resolution of the time-dependent \Sch equation
\beq
i\hbar \frac{\partial}{\partial t}\Psi(\ve{r},b,t)=
\left\{H_0 + U_{cT}\left[\ve{R}_{c}(t)\right]+U_{fT}\left[\ve{R}_{f}(t)\right]-V_{\rm traj}\left[\ve{R}(t)\right]\right\} \Psi(\ve{r},b,t),
\eeqn{e63}
where $V_{\rm traj}$ is the potential used to generate the trajectories and $b$ is the impact parameter, which characterises each trajectory.
The time dependence of the optical potentials $U_{cT}$ and $U_{fT}$ arises from the time dependence of $\ve{R}$, upon which $\ve{R}_{c}$ and $\ve{R}_{f}$ depend (see \Fig{f7}).

The \Eq{e63} has to be solved for each value of the impact parameter $b$ with the condition that the projectile is initially in its ground state:
\beq
\Psi^{(m_0)}(\ve{r},b,t\rightarrow-\infty)=\phi_{n_0l_0m_0}(\ve{r})\hspace{10mm}\forall b\in \R^+.
\eeqn{e64}

The usual way to solve the time-dependent equation \eq{e63} is to numerically compute the wave function $\Psi$ by small time steps $\Delta t$ using an approximation of the time-evolution operator $U$:
\beq
\Psi^{(m_0)}(\ve{r},b,t+\Delta t)&=&U(t+\Delta t,t)\,\Psi^{(m_0)}(\ve{r},b,t)
\eeqn{e65}
with
\beq
U(t',t)=\exp\left\{-\frac{i}{\hbar}\int_t^{t'}\left[H_0 + U_{cT}(\tau)+U_{fT}(\tau)-V_{\rm traj}(\tau)\right]d\tau\right\},
\eeqn{e66}
starting from $\phi_{n_0l_0m_0}$ at a large negative time $t\rightarrow-\infty$.

Since each trajectory is treated separately, the computational cost of the TD approach is strongly reduced compared to CDCC's.
However, as we will see in \Sec{benchmark}, because of that some quantal effects are lost, in particular the model lacks the interferences between neighbouring trajectories.

Various codes---mostly unpublished---have been written to solve \Eq{e63}.
In Refs.~\cite{KYS94,EBB95,TW99}, the wave function $\Psi$ is expanded in partial waves and the $P$-$T$ optical potentials are decomposed into multipoles.
This technique has the advantage to lead to a diagonal representation of $H_0$, which is simpler to use in the expression of the time-evolution operator $U$.
\Ref{Fal02} makes use of a three-dimensional cubic mesh in $\ve{r}$ upon which $\Psi$ is expanded.
This enables to treat the kinetic-energy term of $H_0$ using a fast Fourier transform.
Finally, in \Ref{CBM03c}, the wave function is expanded on a three-dimensional spherical mesh, which leads to a diagonal representation of the $P$-$T$ optical potentials.


\begin{figure}[t]
\center
\includegraphics[width=9cm]{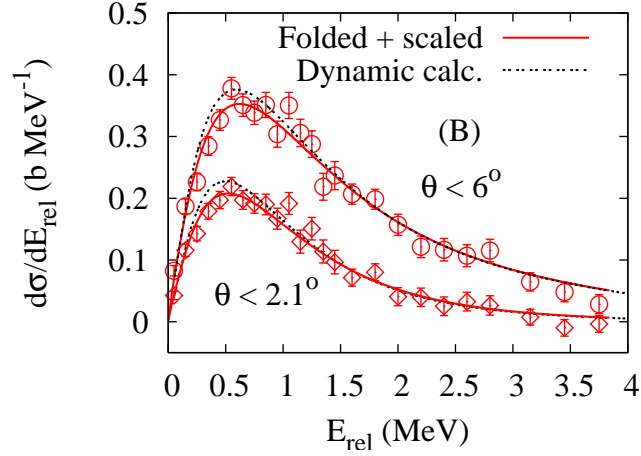}
\caption{\label{f9} Coulomb breakup cross section for $^{15}$C impinging on Pb at $68A$~MeV measured at RIKEN \cite{Nak09} with two angular cutoffs.
The TD calculation of Esbensen is shown with the dotted lines \cite{Esb09,Esb09Err}, the solid lines correspond to these calculations folded with the experimental energy resolution.
Reprinted figure with permission from \Ref{Esb09} Copyright (2009) by the
American Physical Society.}
\end{figure}

\Fig{f9} illustrates this method; it displays the cross section obtained by Esbensen within his TD code for the breakup of $^{15}$C into $^{14}$C and a neutron while impinging on Pb at $68A$MeV \cite{Esb09,Esb09Err}, which correspond to the conditions of the RIKEN experiment performed by Nakamura \etal\ \cite{Nak09}.
Once folded with the experimental resolution (solid red lines in \Fig{f9}), the TD calculations are in excellent agreement with the data, which confirms, besides the validity of the TD approach for this kind of observable, the clear two-cluster structure of $^{15}$C.

\subsubsection{\label{eikonal}Eikonal Approximation}
The second approximation, which is often used to model reactions involving halo nuclei, is the \emph{eikonal} approximation.
This approximation was first derived by Glauber \cite{Glauber} and is suited to describe high-energy reactions.
The cornerstone of this approximation is to realise that at sufficiently high energy, the $P$-$T$ relative motion is not very different from the incoming plane wave $e^{iKZ}$ [see \Eq{e57}].
At high energy, the fragments of the projectile are emitted at a velocity close to the projectile one and are scattered at very forward angles.
The idea of the approximation is thus to factorise the plane wave out of the three-body wave function by posing
\beq
\Psi(\ve{r},\ve{R})=e^{iKZ}\ \widehat{\Psi}(\ve{r},\ve{R}).
\eeqn{e67}
Inserting this in the \Sch equation \eq{e56}, the only term that requires some attention is the kinetic operator $T_R$
\beq
T_R\Psi(\ve{r},\ve{R})=e^{iKZ}\left(T_R +v P_Z +\frac{\mu_{PT}}{2}v^2\right)\widehat{\Psi}(\ve{r},\ve{R}),
\eeqn{e68}
where $v=\hbar K/\mu_{PT}$ is the asymptotic $P$-$T$ relative velocity.
Since the major dependence on $\ve{R}$ has been removed from the wave function through the factorisation \eq{e67}, $\widehat{\Psi}$ is smoothly varying with $\ve{R}$ and hence its second-order derivatives $T_R\widehat{\Psi}$ can be neglected compared to its first-order derivative $vP_Z\widehat{\Psi}$.
Accounting for the energy conservation $E_T=\mu_{PT}v^2/2+E_{n_0l_0}$, \Eq{e56} then reads
\beq
i\hbar v \frac{\partial}{\partial Z}\widehat{\Psi}(\ve{r},\ve{b},Z)
=\left[H_0-E_{n_0l_0}+U_{cT}(\ve{R}_{c})+U_{fT}(\ve{R}_{f})\right]
\widehat{\Psi}(\ve{r},\ve{b},Z),
\eeqn{e69}
where $\ve{b}$ is now the transverse component of $\ve{R}$ (see \Fig{f7}), not to be mixed up with the classical impact parameter of the previous section.
Note that $\ve{b}$ is still a quantal variable here; unlike in the TD approach, no semiclassical hypothesis has been made in the derivation of \Eq{e69}.

Following the initial condition \eq{e57}, \Eq{e69} has to be solved knowing that
\beq
\widehat\Psi^{(m_0)}(\ve{r},\ve{b},Z)\flim{Z}{-\infty}\phi_{n_0l_0m_0}(\ve{r})\hspace{10mm}\forall\ve{b}\in \R^2.
\eeqn{e70}
At this level of approximation, we obtain the \emph{Dynamical Eikonal Approximation} (DEA) \cite{BCG05,GBC06}.
For completeness, note that a similar approximation called the \emph{Eikonal-CDCC} (E-CDCC) has been derived by the Kyushu group \cite{OYI03}.
It is based on the CDCC expansion but exploits the eikonal approximation to simplify the coupled equations \eq{e60}.

Note that \Eq{e69} and its initial condition \eq{e70} are mathematically equivalent to Eqs.~\eq{e63} and \eq{e64}, respectively, which means that it can be solved using the same codes as that time-dependent \Sch equation.
The DEA has thus a computational cost similar to TD calculations, hence significantly lower than CDCC.
The major difference with the TD approach is that $\ve{b}$ being a quantal variable, interference effects between neighbouring trajectories can be taken into account.
The DEA hence extends the TD technique by including part of the quantal interferences that are neglected within the semiclassical approximation.
We will come back to that in the next section.

What people usually call the eikonal approximation performs a subsequent \emph{adiabatic}---or \emph{sudden}---approximation to simplify \Eq{e69}.
Since the factorisation \eq{e67} is applied at high energy, the idea of this additional approximation is to say that the collision time will be short and that during this time, the projectile structure will not evolve much.
One can thus see the internal coordinates of the projectile, i.e. $\ve{r}$, as being frozen during the collision.
This amounts to neglect the influence of the projectile Hamiltonian and set
\beq
H_0\sim E_{n_0l_0}.
\eeqn{e71}
With that adiabatic approximation, the solutions of \Eq{e69} that corresponds to the incoming condition \eq{e70} can be easily computed and read
\beq
\widehat{\Psi}^{(m_0)}_{\rm{eik}}(\ve{r},\ve{b},Z)
=\exp\left\{-\frac{i}{\hbar v}\int_{-\infty}^{Z}
\left[U_{cT}(\ve{r},\ve{b},Z')+U_{fT}(\ve{r},\ve{b},Z')\right]dZ'\right\} \phi_{n_0l_0m_0}(\ve{r}).\hspace{4mm}
\eeqn{e72}
This expression can be easily interpreted: the projectile is seen as following a straight-line trajectory ($\ve{R}=\ve{b}+Z\ve{\widehat{Z}}$) along which it accumulates a phase through its interaction with the projectile.
As can be seen from expression \eq{e72}, this model of the reaction is quite simple to implement and has a significantly shorter computational cost compared to CDCC, and even to TD methods.
The fact that the eikonal wave function \eq{e72} includes the initial ground state of the projectile $\phi_{n_0l_0m_0}$ translates the fact that the internal coordinate $\ve{r}$ of the projectile is seen as frozen during the collision.

\begin{figure}[t]
\center
\includegraphics[width=9cm]{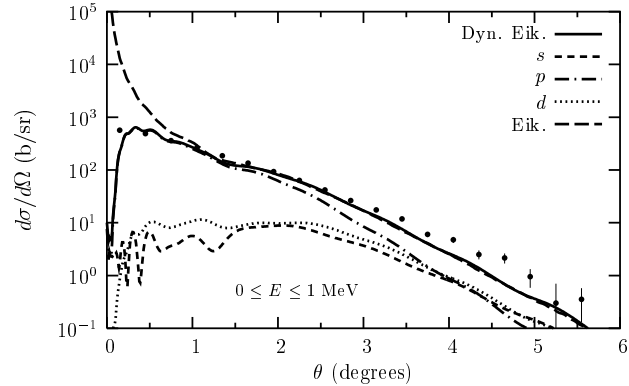}
\caption{\label{f10} Angular distribution for the Coulomb breakup of $^{11}$Be on Pb at $69A$~MeV measured at RIKEN \cite{Fuk04}.
The DEA calculation is in excellent agreement with the data.
The usual eikonal approximation diverges at forward angle due to its additional adiabatic approximation, which is invalid for the infinitely ranged Coulomb force \Ref{GBC06}.
Reprinted figure with permission from \Ref{GBC06} Copyright (2006) by the
American Physical Society.}
\end{figure}

This approximation will provide accurate results only if the adiabatic approximation is well justified.
If the dynamics of the projectile plays a significant role, then the DEA or the E-CDCC have to be considered.
This is the case for Coulomb-dominated reactions.
The long---actually infinite---range of the Coulomb interaction makes the sudden approximation invalid. 
In that case, the collision time cannot be considered as brief, since the projectile will never end interacting with the Coulomb field of the target and hence couplings within the continuum, like the post-acceleration of the core, must be accounted for in the description of the reaction.
This is illustrated in \Fig{f10}, where the breakup cross section for $^{11}$Be impinging on Pb at $69A$~MeV is plotted as a function of the scattering angle $\theta$ of the centre of mass of the $^{10}$Be and n constituents after dissociation with a relative energy $0\le E\le1$~MeV \cite{GBC06}.
The DEA calculation (solid line) is in excellent agreement with the data of RIKEN \cite{Fuk04}.
The contributions of the main partial waves to the cross section ($s$ wave with short-dashed line, $p$ wave with dash-dotted line and $d$ waves with dotted line) show that at forward angle the process is dominated by a transition towards the $^{10}$Be-n continuum in the $p$ wave, i.e. a direct E1 transition from the $s$ ground state of $^{11}$Be.
The usual eikonal approximation (long-dashed line), however, diverges at forward angles, confirming that the adiabatic approximation does not hold in this Coulomb case.
This result emphasises the importance to know the range of validity of the reaction model one uses when analysing experimental data.

\subsection{\label{benchmark}Benchmark of Breakup Models}
Various comparisons between reaction models have been performed recently \cite{CEN12,UDN12}.
In this section, I present the results of \Ref{CEN12}, where the three models presented in the previous section, CDCC, TD and DEA, have been compared to one another on the test case of the breakup of $^{15}$C on Pb at $68A$~MeV, which has been measured at RIKEN \cite{Nak09}.

\begin{figure}[t]
\center
\includegraphics[width=5.8cm]{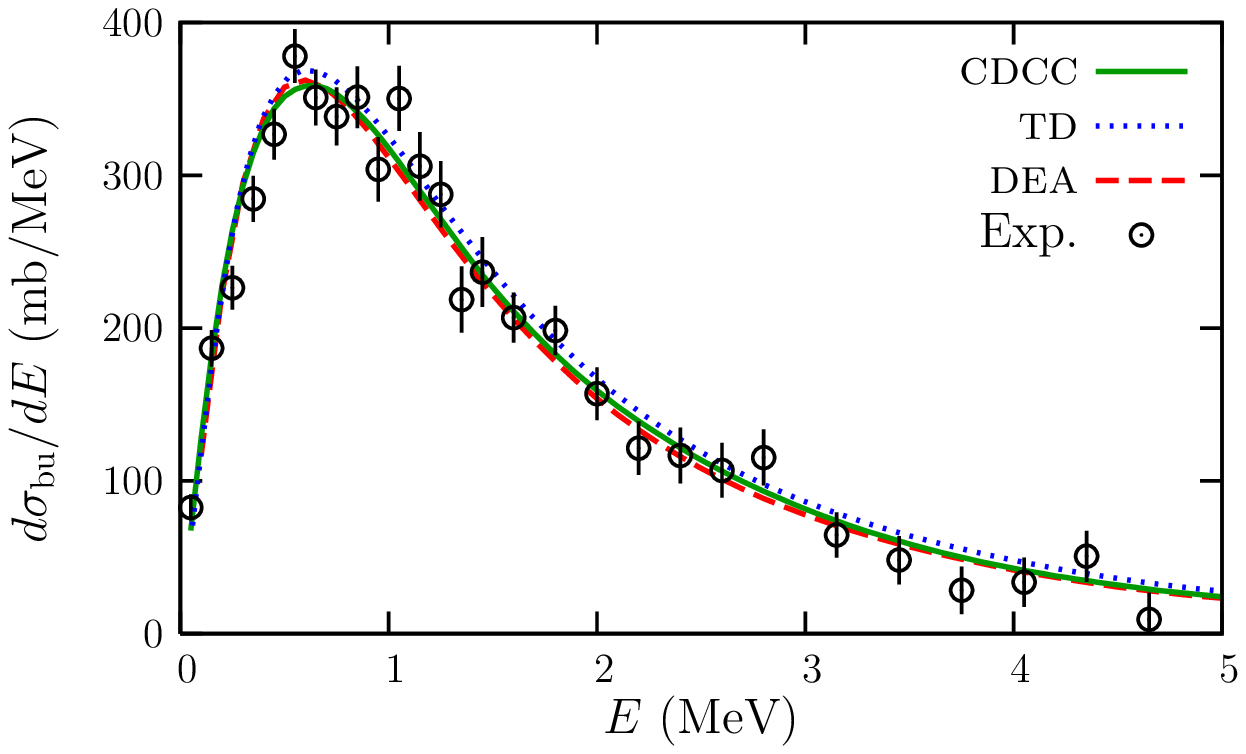}
\includegraphics[width=5.8cm]{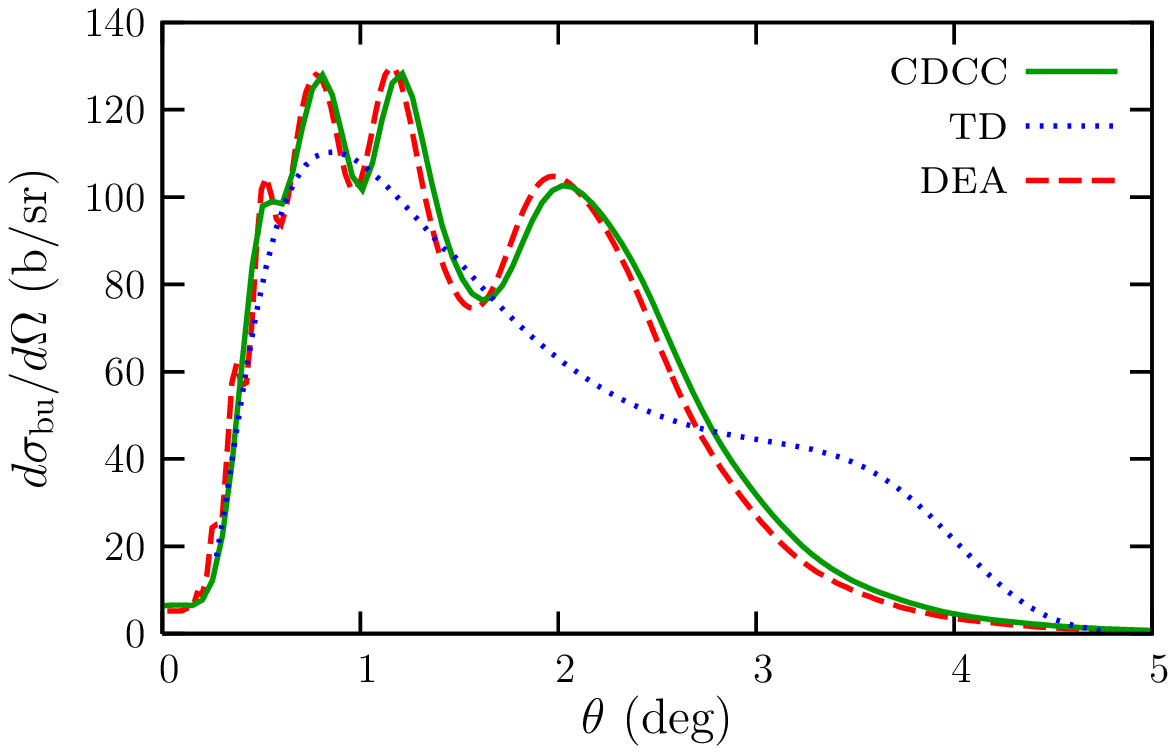}
\caption{\label{f11} Comparison of breakup models (CDCC, TD and DEA) on the Coulomb breakup of $^{15}$C on Pb at $68A$~MeV measured at RIKEN \cite{Nak09}.
Left: energy distribution. Right: angular distribution \cite{CEN12}.
Reprinted figures with permission from \Ref{CEN12} Copyright (2012) by the
American Physical Society.}
\end{figure}

Using the same $^{14}$C-n potential $V_{cf}$ and the same optical potentials $U_{cT}$ and $U_{fT}$, the breakup cross section has been computed within each of the reaction models.
The results are summarised in \Fig{f11}.
The energy distribution displayed in the left panel shows that all three models provide nearly identical cross sections when the same two-body potentials are considered in input.
This shows that, at this beam energy and for this observable, all three models are equivalent.
Incidentally, their predictions are in excellent agreement with the RIKEN data \cite{Nak09}, which confirms the two-body structure of $^{15}$C.

The differences appear in the right panel of \Fig{f11}, which presents the breakup cross section as a function of the scattering angle $\theta$ of the $^{14}$C-n centre of mass after dissociation.
For that observable, we observe oscillations in both the CDCC (solid line) and DEA (dashed line) cross sections, whereas the TD calculation (dotted line) provides a smooth observable.
This is the sign of the missing quantal effects within the semi-classical approximation mentioned in \Sec{TD}.
Both CDCC and the DEA includes these interferences.
They also show excellent agreement with one another at this energy, although the DEA is much less time consuming than CDCC.
Note however, that the TD model provides a cross section that follows the general trend of the quantal results.
This explains that the agreement of the TD energy distribution with the other two observed in \Fig{f11}~(left) is not accidental.
Since that observable is obtained after integration over the scattering angle $\theta$, the oscillations around the general trend cancel out, which leads to identical energy distributions.

In \Ref{CEN12}, the comparison has also been extended to lower beam energy, viz. $20A$~MeV, where the eikonal approximation is no longer supposed to provide a reliable description of the reaction.
As expected, it is observed that the DEA is no longer in agreement with CDCC: the DEA breakup cross section is too large and focused at too forward a scattering angle compared to CDCC's.
This comes from the hypothesis that the $P$-$T$ relative motion does not differ much from the incoming plane wave (see \Sec{eikonal}).
The semi-classical interpretation of this eikonal approximation, in which the projectile follows a straight-line trajectory, suggests that the projectile is then forced to pass through the high-field zone of the target, in which it would otherwise not enter at low beam energy due to the significant Coulomb repulsion by the target.
The TD approximation, which naturally includes the $P$-$T$ repulsion through the semi-classical (non-linear) trajectory, does not exhibit such a flaw and hence provides an agreement with CDCC similar to what is seen in \Fig{f11}: a nearly identical energy distribution and an angular distribution which lacks the quantal oscillatory pattern but otherwise follows the trend of the CDCC prediction \cite{CEN12}.

To account for the $P$-$T$ Coulomb repulsion within the DEA (or E-CDCC), it has been suggested to use a semi-classical correction, in which the norm of the transverse part $\ve{b}$ of the $P$-$T$ relative coordinate $\ve{R}$ is replaced by the distance of closest approach in the corresponding Coulomb trajectory \cite{BD04}.
With that correction, the DEA and E-CDCC calculations fall in perfect agreement with the CDCC prediction, including the effect of the quantal interferences \cite{FOC14}.
In this way, at least for Coulomb-dominated reactions, the eikonal approximation in its most general expression, i.e. without the adiabatic approximation, can be safely extended to low energies.
Since this model of reactions is significantly less time consuming than CDCC, this is not a vain gain.

\section{\label{structure}Application of Breakup Reactions to Nuclear-Structure Study}

\subsection{Binding Energy}
\begin{figure}[th]
\sidecaption[t]
\includegraphics[trim=1.8cm 0.9cm 2.3cm 11cm, clip=true, width=7.5cm]{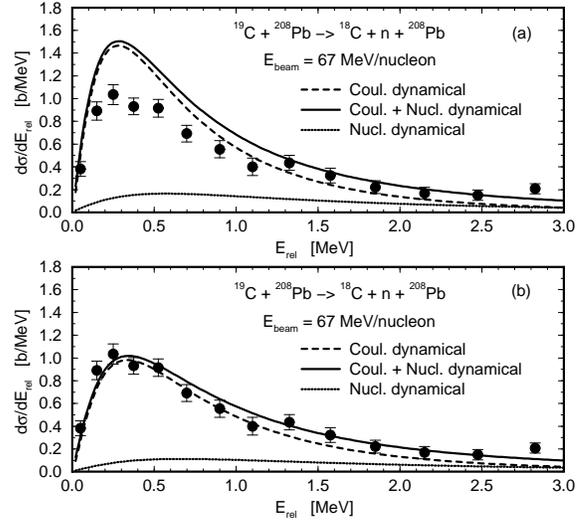}
\caption{\label{f12} Coulomb breakup of $^{19}$C on Pb at $67A$~MeV measured at RIKEN \cite{Nak99}.
The theoretical analysis of Typel and Shyam illustrates the influence of the binding energy of the system on the breakup energy distribution \cite{TS01}.
$S_{\rm n}(^{19}{\rm C})=530$~keV (top) and 650~keV (bottom).
Reprinted figures with permission from \Ref{TS01} Copyright (2001) by the
American Physical Society.}
\end{figure}

The first observable to which breakup reactions are sensitive is the binding energy of the projectile.
This is illustrated in \Fig{f12}, where the Coulomb-breakup cross section of $^{19}$C---another candidate one-neutron halo nucleus---is displayed as a function of the $^{18}$C-n energy $E$ after dissociation.
The beam energy is $67A$~MeV and the target is Pb.
The experiment was performed at RIKEN \cite{Nak99} and the theoretical analysis illustrated in \Fig{f12} is due to Typel and Shyam using a TD approach (see \Sec{TD}) \cite{TS01}.
In this analysis, $^{19}$C is described as an inert $^{18}$C core to which a $1s_{1/2}$ neutron is loosely bound.
At the time, the binding energy of $^{19}$C was poorly known, which is why two series of calculations were made, corresponding to two realistic choices of the one-neutron separation energy for $^{19}$C: $S_{\rm n}(^{19}{\rm C})=530$~keV (top panel) and 650~keV (bottom panel).

Typel and Shyam's analysis clearly shows that the breakup cross section is significantly sensitive to the binding energy of the system.
This can be easily understood at a qualitative level: the less a system is bound the easier it will be to break.
We indeed see that calculations for a $^{19}$C bound by 530~keV (top panel of \Fig{f12}), the breakup cross section is about 50\% higher than when it is bound by 650~keV.
In addition to the change in magnitude, we also witness a variation in the dependence of the cross section on the continuum energy $E$.
The RIKEN data are better reproduced in both magnitude and shape when $S_{\rm n}(^{19}{\rm C})$ is set to 650~keV.
Since this analysis, the mass of $^{19}$C has been accurately measured and its binding energy has been precisely determined to be $580\pm9$~keV \cite{NNDC}.
This value is clearly higher than 530~keV, but it is lower than the suggested value of Typel and Shyam of 650~keV.

In addition to its sensitivity to the binding energy of $^{19}$C, the analysis of Typel and Shyam also shows how the breakup cross section is affected by the Coulomb and nuclear parts of the interaction between the projectile and the target.
The thick solid (higher) lines in \Fig{f12} correspond to the full calculation, i.e. containing both the Coulomb and nuclear parts of the $P$-$T$ interaction.
Purely Coulomb---yet fully dynamical---calculations are shown by the dashed lines.
The cross sections obtained with the sole nuclear interaction are displayed by the thin solid (lower) lines.
As expected, this reaction is strongly dominated by the Coulomb interaction.
Yet, the nuclear part of the interaction has non-negligible effects, especially at high $^{18}$C-n energy.

A systematic analysis of the sensitivity of the Coulomb-breakup cross section to the projectile structure (binding energy, orbital angular momentum in the ground state $l_0$,\ldots) has been made by Typel and Baur within a perturbative solution of the TD approach assuming a purely Coulomb $P$-$T$ interaction \cite{TB04,TB05}.
This seminal work provides an analytical expression for the cross section.
Although the influence of the nuclear interaction is entirely neglected, and the effects of orders beyond the first one, i.e. beyond a mere one-step transition between the ground state and the continuum, this enables them to explain in simple terms how the shape of the energy distribution depends on the structure of the projectile.
This helps identifying the key degrees of freedom to properly describe Coulomb breakup.

\subsection{\label{spectro}Spectroscopy}
All the models of breakup presented in \Sec{bu} are based on a single-particle description of the projectile [see \Eq{e54}], where the internal structure of the core is neglected and the bound states of the projectile are described within a single configuration with a wave function of norm 1.
In reality the structure of any nucleus $^AP$ in its state of spin and parity $J^\pi$ is an admixture of various configurations in which the core $^{A-1}c$ can be in different states $ci$ of spin and parity $J_{ci}^{\pi_{ci}}$ (see A.~Brown lectures to this summer school)
\beq
^AP(J^\pi)= \sum_{ci}\ \left[^{A-1}c(J_{ci}^{\pi_{ci}})\otimes \psi^{ci}_{l}\right]^{JM}
\eeqn{e73}
For each configuration, the \emph{overlap wave function} $\psi^{ci}_{lm}$ describes the relative motion of the halo neutron to the core in a given state $ci$.
The square of the norm of this overlap wave function measures the probability to find the system in the configuration $ci$.
It is called the \emph{spectroscopic factor} (SF)
\beq
{\cal S}^{ci}_{l}=\|\psi^{ci}_{lm}\|^2.
\eeqn{e74}
The structure of halo nuclei is usually dominated by one configuration $c0$, which provides the largest contribution to the breakup cross section.
It has therefore been suggested to extract spectroscopic factors from the comparison of reaction calculations to actual breakup data.
This idea is based on the approximation that the overlap wave function $\psi^{c0}_{l_0m_0}$ can be well approximated by the single-particle wave function $\phi_{n_0l_0m_0}$ \eq{e55}
\beq
\psi^{c0}_{l_0m_0}(\ve{r})=\sqrt{{\cal S}^{c0}_{l_0}}\ \phi_{n_0l_0m_0}(\ve{r}).
\eeqn{e75}
This single-particle approximation leads to the idea that the spectroscopic factor can be obtained through the simple ratio
\beq
{\cal S}^{c0}_{l_0}=\frac{\sigma_{\rm bu}^{\rm exp}}{\sigma_{\rm bu}^{\rm th}},
\eeqn{e76}
where the experimental cross section $\sigma_{\rm bu}^{\rm exp}$ is compared to the theoretical prediction $\sigma_{\rm bu}^{\rm th}$ starting from the initial single-particle ground-state wave function $\phi_{n_0l_0m_0}$.

The question I would like to raise here is whether this approach is valid for breakup.
Because of the large extension of the halo, this reaction is expected to be rather \emph{peripheral}, in the sense that it should probe mostly the halo structure, i.e. the tail of the projectile wave function.
If this is the case, we would expect the breakup cross section to scale with the normalisation of the tail of the $c$-$f$ wave function [see \Fig{f6} (right)] rather than with the norm of the whole overlap wave function, i.e. the spectroscopic factor ${\cal S}^{c0}_{l_0}$.
At large distance, i.e. beyond the centrifugal barrier, the effect of the short-range nuclear $c$-$f$ potential is negligible and the behaviour of the radial wave function is exactly known, but for its normalisation.
For the effective single-particle wave function $\phi_{n_0l_0m_0}$ it reads (the radial part of the actual overlap wave function $\psi^{c0}_{l_0m_0}$ has an identical behaviour)
\beq
u_{\kappa_{n_0l_0}l_0}(r)\flim{r}{\infty} {\cal C}_{n_0l_0}\ e^{-\kappa_{n_0l_0} r},
\eeqn{e77}
where the wave number $\kappa_{n_0l_0}$ is related to the projectile binding energy $E_{n_0l_0}=-\hbar^2\kappa_{n_0l_0}^2/2\mu_{cf}$, and ${\cal C}_{n_0l_0}$ is the Asymptotic Normalisation Coefficient (ANC), which measures the probability strength in the halo.
The value of this ANC depends on the short-range physics, i.e. the particulars of the effective $V_{cf}$ potential in the single-particle viewpoint.

To test which part of the wave function is actually probed, breakup calculations have been performed using two choices for $V_{cf}$ that lead to radial single-particle wave functions with identical asymptotics, i.e. identical ANCs, but with significant different interiors \cite{CN07}.
This was achieved using supersymmetric transformations of deep $V_{cf}$ potentials \cite{Bay87l,Bay87}.
The deep potentials are chosen to host, in addition to the physical loosely bound state, a deeply spurious bound state in the ground-state partial wave of orbital angular momentum $l_0$.
The presence of that state adds a node in the radial wave function of the physical loosely bound state.
Through supersymmetric transformations the spurious deeply bound state can be removed without affecting the long-range physics of the projectile Hamiltonian $H_0$, viz. the ANC of the other bound states and the phaseshift in the continuum.

\begin{figure}[th]
\center
\includegraphics[width=5.8cm]{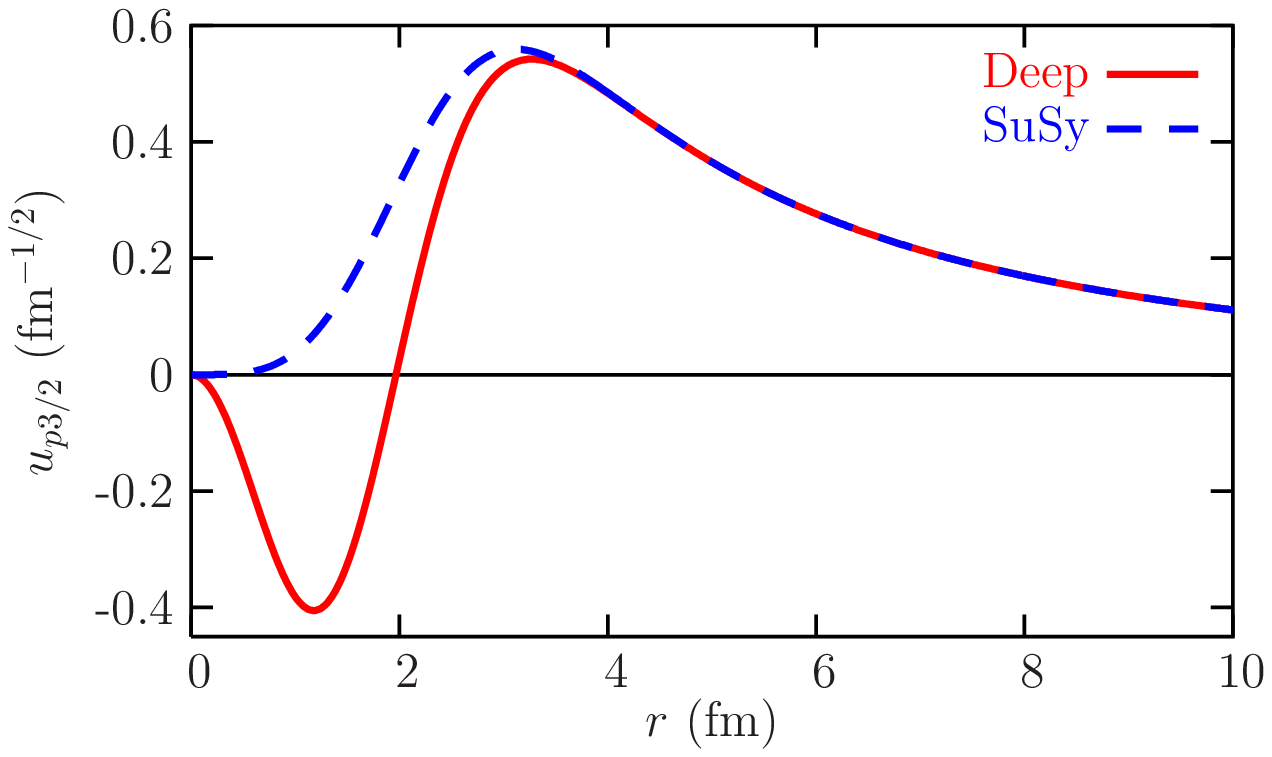}
\includegraphics[trim=2cm 18.3cm 6.5cm 2cm, clip=true, width=5.8cm]{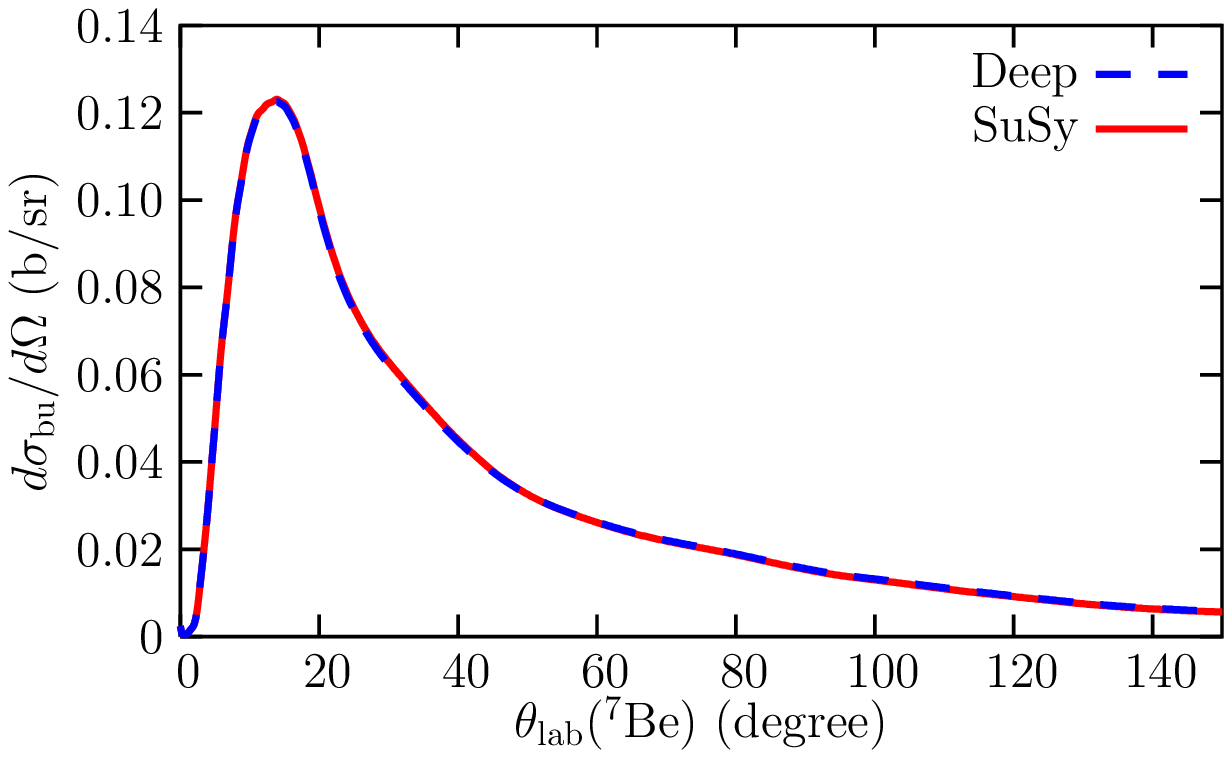}
\caption{\label{f13} Influence of the internal part of the projectile wave function upon breakup calculations \cite{CN07}.
Using two descriptions of $^8$B with identical asymptotics but strongly different interiors (left), the breakup calculations provide identical cross sections, shown here on the right as an angular distribution obtained on a nickel target at 26~MeV.
Reprinted figures with permission from \Ref{CN07} Copyright (2007) by the
American Physical Society.}
\end{figure}

This provides us with two descriptions of the projectile with identical ANCs but strongly different interiors: one with a  node and one without a node.
This is illustrated in \Fig{f13} (left) in the case of $^8$B, the best known one-proton halo candidate, which is thus described as a $^7$Be core to which a valence proton is loosely bound.
The $^7$Be-p radial wave function obtained with the deep potential exhibits a node at $r\simeq2$~fm (solid red line), which disappears once the spurious deeply bound state is removed by the sypersymmetric transformations (blue dashed line).
Note that by construction, both wave functions exhibit exactly the same tail, i.e. the same ANC.

If the reaction process is sensitive to the internal part of the projectile wave function, significant differences should appear in the breakup cross section.
However, the calculations performed with such pair of descriptions show no sensitivity to the choice of the potential.
This is illustrated in \Fig{f13} (right) in the particular case of the angular distribution for the breakup of $^8$B on Ni at 26~MeV, which correspond to the conditions of the Notre-Dame experiment \cite{Kol01}.
The calculations were performed within the CDCC framework as in \Ref{TNT01}.
Although both descriptions of $^8$B exhibit significant differences [see \Fig{f13} (left)], the corresponding breakup cross sections are superimposed on one another.
This result is very general as it is observed for both Coulomb- and nuclear-dominated reactions, at low and high beam energies, for one-proton and one-neutron halo nuclei, and is valid for various kinds of breakup observables (energy and angular distributions) \cite{CN07}.
This clearly shows that the reaction process is purely peripheral and hence probes only the ANC of the initial bound state.
Since the reaction process is not sensitive to the internal part of the wave function, it cannot be sensitive to the norm of the whole wave function.
The spectroscopic factors extracted from such measurements are thus highly questionable.

\subsection{\label{resbu}Resonances}
\begin{figure}[th]
\center
\includegraphics[trim=4cm 15.5cm 4cm 4cm, clip=true, width=8cm]{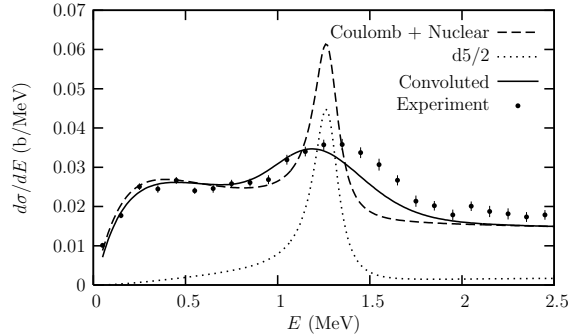}
\caption{\label{f14} Influence of a resonance within the low-energy $^{10}$Be-n continuum upon the breakup of $^{11}$Be on C at $67A$~MeV \cite{CGB04,Fuk04}.
The contribution of the $d_{5/2}$ partial wave, which hosts the resonance, is responsible for the sharp peak in the theoretical cross section.
Once folded with the experimental energy resolution, the results of this TD calculation fit nicely with the data of \Ref{Fuk04}.}
\end{figure}

In addition to its two bound states, the $^{11}$Be spectrum hosts a resonance at low energy in its continuum.
It is a $\fial^+$ resonance, which is interpreted as a single-particle resonance in the $d_{5/2}$ partial wave \cite{CGB04} (see \Sec{resonances}).
Measuring the breakup of $^{11}$Be on carbon at $67A$~MeV, Fukuda \etal\ have observed a large peak in the $^{10}$Be-n continuum at the energy of this resonance \cite{Fuk04}, see \Fig{f14}.

Within a TD technique, it has been confirmed that the increase of the breakup strength in that region could be explained when the $^{10}$Be-n potential is fitted to host such a resonance in the $d_{5/2}$ partial wave \cite{CGB04}, see \Fig{f14}.
The sharp peak observed in the theoretical cross section (dashed line) is entirely due to the contribution of the $d_{5/2}$ partial wave that hosts the resonance (dotted line).
Interestingly, the width of that peak matches that of the resonance, confirming that nuclear-dominated breakup can provide significant information about structures within the continuum.
Once folded with the experimental energy resolution, these TD calculations come quite close to the data of \Ref{Fuk04}.

\subsection{Role of the Non-Resonant Continuum}
In the previous section, we have seen that resonant structures within the continuum could affect nuclear-dominated breakup reactions.
In this section, let us explore the sensitivity of breakup calculations to the non-resonant part of the continuum.
One way to do so is to fit various $c$-$f$ potentials, e.g. with different geometries, to the same nuclear-structure inputs.
This was done, for example in \Ref{CN06}, where different $^{10}$Be-n potentials of Woods-Saxon shape were fitted to reproduce the $\half^+$ ground state of $^{11}$Be within the $1s_{1/2}$ orbit, its first excited state $\half^-$ in the $0p_{1/2}$, and the $\fial^+$ resonance in the $d_{5/2}$ partial wave.

\begin{figure}[b]
\center
\includegraphics[width=8.5cm]{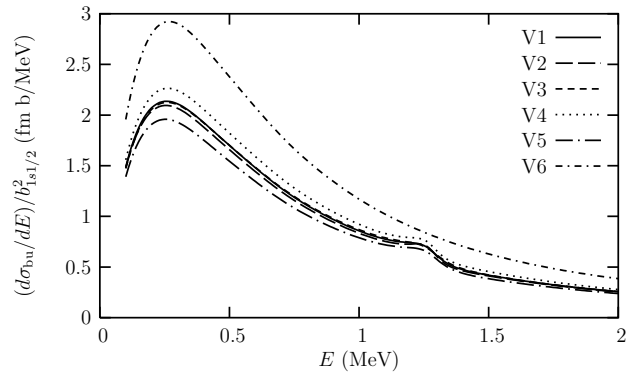}
\caption{\label{f15} Influence of the non-resonant continuum upon the breakup cross section.
Example on the Coulomb breakup of $^{11}$Be on Pb at $69A$~MeV \cite{CN06}.
The sensitivity to the ground state ANC has been removed by dividing the cross section by the square of the ANC.
The remaining difference comes mostly from the differences in the $p_{3/2}$ partial wave, which has not been constrained in the construction of the different $^{10}$Be-n potentials.
Reprinted figure with permission from \Ref{CN06} Copyright (2006) by the
American Physical Society.}
\end{figure}

The breakup cross section on Pb at $69A$~MeV computed with these potentials within a TD approach are shown in \Fig{f15} \cite{CN06}.
To remove the significant dependence on the ANC mentioned in \Sec{spectro}, they have been divided by the square of the ANC of the single-particle wave function of the ground state (denoted in \Fig{f15} by $b_{1s1/2}$)
We observe that although the main dependence on the ANC has been removed, there remains a large variation in the breakup cross sections obtained with the different potentials.
A detailed analysis shows that most of that sensitivity comes from the $p_{3/2}$ partial wave, which is not constrained by any physical observable.
The variation between the $p_{3/2}$ contributions to the breakup cross sections obtained with the$^{10}$Be-n potentials built in \Ref{CN06} can be traced back to the changes induced in the $p_{3/2}$ phaseshifts.

Such an influence of the non-resonant continuum upon the breakup cross section is very general.
Similar results have also been observed for the breakup of a $^8$B projectile at low energy \cite{CN06}.
This effect should be taken into account in the analysis of experiments, as it can significantly affect theoretical predictions.
In particular, this could spoil the extraction of an ANC for the initial ground state from the direct---and naive---comparison of calculations to data.
A similar conclusion has been drawn in a recent work using a Halo-EFT description of the projectile \cite{CPH18}.

\subsection{Effect of Core Excitation}
So far, we have seen models in which the core was seen as a structureless body, of which the spin and parity are neglected.
Although this is usually a good approximation, especially when the ground state of the core is a $0^+$ state and the energy of its first excited state is large, other configurations are possible, as mentioned in \Sec{spectro}.
Including these other configurations in reaction models is a tricky business, not only because it increases the complexity of the structure of the projectile, but also because the reaction mechanism must then include the possible dynamical excitation of the core, i.e. reaction channels in which the core is excited by the target during the collision.

The first attempt to include the core excitation within the CDCC framework was performed by Summers \etal\ \cite{SNT06R,SNT06,SNT06Err}.
It led to the development of the XCDCC model, which stands for eXtended CDCC.
However, the effect of this core excitation in the case of $^{11}$Be seemed then rather small \cite{SN07,SN07Err}.

\begin{figure}[th]
\sidecaption
\includegraphics[trim=0.3cm 1.1cm 3.2cm 4.5cm, clip=true, width=6cm]{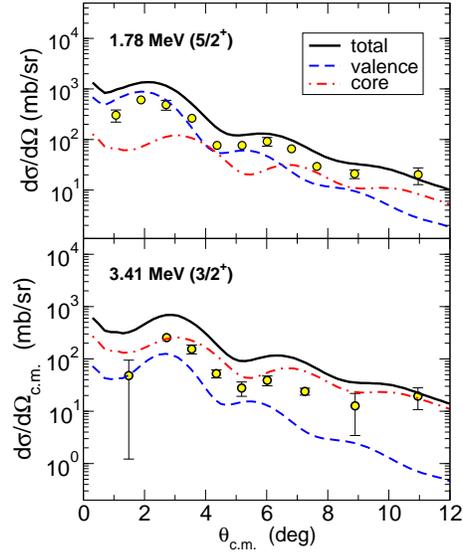}
\caption{\label{f16} Influence of the core excitation on the breakup of $^{11}$Be on C at $67A$~MeV \cite{ML12}.
The shape of the data for these angular distributions can be reproduced only if the core excitation is included in these DWBA calculations.
Data are from \Ref{Fuk04}.
Reprinted figure with permission from \Ref{ML12} Copyright (2012) by the
American Physical Society.}
\end{figure}

More recently, Moro and Lay have extended a DWBA code to include the core excitation \cite{ML12}.
This enabled them to study the resonant breakup of $^{11}$Be on C at $67A$~MeV, already mentioned in \Sec{resbu} .
Their results are illustrated in \Fig{f16}, which displays the breakup cross sections as angular distributions at the energy of the $\fial^+$ (top) and $\thal^+$ (bottom) resonances within the $^{11}$Be continuum.
Their calculations (black solid lines) are in very good agreement with the data.
In particular, the angular dependence of the cross sections perfectly matches that of the RIKEN data \cite{Fuk04}.
Such a good result is obtained only if both the excitation of the valence neutron to the continuum (red dash-dotted lines) and the core excitation (blue dashed lines) are included together in the reaction model.
For the $\fial^+$ resonance (\Fig{f16} top), we see that albeit small, the effect of the core excitation is required to obtain an angular dependence that fits the data.
For the $\thal^+$ state, however, the core excitation is the dominant process in the reaction, suggesting that the structure of this resonance is dominated by a configuration in which the $^{10}$Be core is in its $2^+$ excited state.

These results indicate that for some observables, the core excitation can be a significant part of the breakup process, especially at higher energy in the continuum.
More efforts should therefore be made to include it within models to improve the description of the collision and increase our understanding of reactions involving exotic nuclei.

\section*{Conclusion}
Reactions are used in many fields of quantum physics.
Whether in the field of molecular, atomic, nuclear, or particle physics, significant amounts of information can be gathered through the study of collisions.
The present notes offer an introduction to quantum scattering theory explained in the realm of nuclear physics.
The extension of that theory to reactions in which the projectile can break up into two more elementary constituents and its particular application to the study of halo nuclei has also been presented.
The three major models of breakup reaction have been detailed, the CDCC method, the TD approach and the eikonal approximation.
A benchmark of these models on the special case of $^{15}$C shows the limitation of each model, which enables us to estimate the range of validity of each of these approaches.

In the last section, an analysis of the nuclear-structure information about halo nuclei one can get from breakup measurements has been provided.
Through examples chosen from the literature, the complexity of the reaction mechanism and its sensitivity to the structure of the projectile has been presented.
By showing which part of the projectile wave function is probed during breakup, these results emphasise the importance to study in detail to which structure observable the reaction is actually sensitive.
The blind application of an accepted recipe can lead to misinterpretation of experimental results.

Two key points should be taken away from this brief summary: the first is to pay attention to the domain of validity of the reaction model used to analyse data.
Does the model fit the experimental conditions under which they have been gathered?
For example, it does not make sense to use an eikonal model of reaction to analyse an experiment performed at the Coulomb barrier.
The second is to know to what the reaction is sensitive.
For example, it does not make sense to extract a spectroscopic factor from a peripheral experiment.

With this short review, I hope to have delivered a basic introduction to quantum-reaction modelling within the realm of nuclear physics.
Hopefully, it will have triggered the interest of the reader in reaction theory and provide him/her with a list of useful references to deepen his/her knowledge in this exciting field of physics.

\section*{Acknowledgements}
This project has received funding from the European Union’s Horizon 2020 research and innovation program under grant agreement No 654002, the Deutsche Forschungsgemeinschaft within the Collaborative Research Centers 1245 and 1044, and the PRISMA (Precision Physics, Fundamental Interactions and Structure of Matter) Cluster of Excellence. I also acknowledges the support of the State of Rhineland-Palatinate.


\end{document}